\documentclass[useAMS,usenatbib]{mn2e}

\usepackage{hyphenat}
\usepackage{graphicx}
\usepackage{fixltx2e}
\usepackage{color}
\usepackage{amsmath}
\usepackage{amsfonts}
\usepackage{aas_macros}
\usepackage{dcolumn}
\usepackage{url}
\usepackage{natbib}

\newcommand{\revision}[1]{#1}
\newcommand{\ion}[2]{#1\textsc{#2}}

\newcommand{\vecvar}[1]{\textbf{#1}}
\newcommand{\nn}{{N\!N}}

\DeclareMathOperator{\asinh}{asinh}
\DeclareMathOperator*{\argmin}{arg\,min}

\newcolumntype{d}{D{.}{.}{-1} } 

\title
[Photometric redshifts for the SDSS DR12]
{Photometric redshifts for the SDSS Data Release 12}
\author[R. Beck, L. Dobos, T. Budav\'ari, A.S.~Szalay and I. Csabai]{R\'obert Beck$^{1,2}$\thanks{E-mail:
beckrob23@caesar.elte.hu, dobos@complex.elte.hu, csabai@complex.elte.hu}, L\'aszl\'o Dobos$^{1}$, Tam\'as Budav\'ari$^{3,4,2}$,
\newauthor
Alexander S. Szalay$^{2}$ and Istv\'an Csabai$^{1}$ \\ 
$^{1}$Department of Physics of Complex Systems, E\"{o}tv\"{o}s Lor\'{a}nd University, 1117 Budapest, Hungary \\
$^{2}$Department of Physics and Astronomy, The Johns Hopkins University, Baltimore, MD 21218, USA \\
$^{3}$Department of Applied Mathematics and Statistics, The Johns Hopkins University, Baltimore, MD 21218, USA \\
$^{4}$Department of Computer Science, The Johns Hopkins University, Baltimore, MD 21218, USA}
\begin{document}

\date{Accepted 2016 Month Day. Received 2016 Month Day; in original form 2016 Month Day}

\pagerange{\pageref{firstpage}--\pageref{lastpage}} \pubyear{2016}

\maketitle

\label{firstpage}

\begin{abstract}

We present the methodology and data behind the photometric redshift database of the Sloan Digital Sky Survey Data Release 12 (SDSS DR12). We adopt a hybrid technique, empirically estimating the redshift via local regression on a spectroscopic training set, then fitting a spectrum template to obtain K-corrections and absolute magnitudes. The SDSS spectroscopic catalog was augmented with data from other, publicly available spectroscopic surveys to mitigate target selection effects. The training set is comprised of $1,976,978$ galaxies, and extends up to redshift $z\approx 0.8$, with a useful coverage of up to $z\approx 0.6$. We provide photometric redshifts and realistic error estimates for the $208,474,076$ galaxies of the SDSS primary photometric catalog. We achieve an average bias of $\overline{\Delta z_{\mathrm{norm}}} = \revision{5.84 \times 10^{-5}}$, a standard deviation of $\sigma \left(\Delta z_{\mathrm{norm}}\right)=\revision{0.0205}$, and a $3\sigma$ outlier rate of $P_o=\revision{4.11\%}$ when cross-validating on our training set. The published redshift error estimates and photometric error classes enable the selection of galaxies with high quality photometric redshifts. We also provide a supplementary error map that allows additional, sophisticated filtering of the data.

\end{abstract}


\section{Introduction}
\label{sec:intro}


Photometric redshift estimation has become a vital technique in the field of astronomy, as it enables measuring the distance of a much larger number of objects than what would be achievable through a spectroscopic survey. The Sloan Digital Sky Survey is one of the largest public collections of both photometric and spectroscopic measurements, with $208,478,448$ galaxies in its photometric catalog \citep{York2000, Gunn1998, Doi2010}, and, as of Data Release 12 \citep{Alam2015}, $2,274,081$ galaxy spectra in the continually expanded spectroscopic catalog \citep{Eisenstein2011, Smee2013}.

The purpose of this paper is to give a detailed description of the methods and data we used in creating the photometric redshift database of SDSS DR12, released to the public in January 2015. We chose an empirical technique, local linear regression, to estimate the redshift and its error, utilising a training set of $1,976,978$ elements, assembled from DR12 spectroscopy and data from other spectroscopic surveys (listed in Sec.~\ref{sec:crossmatch}). Additionally, we computed the maximum likelihood spectral template fit to the photometry, using the composite spectrum atlas of \citet{Dobos2012}, to obtain additional information such as K-corrections, spectral type, and rest-frame absolute magnitudes.

The main goal of our photometric redshift catalog is to complement the estimated redshift with a reasonable assessment of the estimation error for the wide variety of galaxies in the SDSS photometric survey. The inclusion of spectroscopic data from other surveys means that we have more reference points for distant and faint bluer objects, up to $z\approx 0.8$ and $r\approx 21.5 \: \mathrm{mag}$, which would be less well-represented in SDSS spectroscopy due to target selection \citep{Eisenstein2001, Dawson2013}. We also published an error map in support of this goal, as it highlights problematic regions in the space of galaxy colours where there are overlapping galaxies at different redshifts, leading to reduced accuracy.

The structure of the paper is as follows. In Sec.~\ref{sec:methods}, we describe our empirical method of redshift estimation, and outline the template fitting procedure. Sec.~\ref{sec:trainingSet} explains how the training set was compiled. Our results are presented and discussed in Sec.~\ref{sec:results}. We give pointers on using the database in Sec.~\ref{sec:database}. We provide a summary in Sec.~\ref{sec:summary}.

Throughout the paper, broad-band magnitudes are quoted in the SDSS $\asinh$ magnitude system \citep{Lupton1999}, and are dereddened according to \citet{Schlegel1998}. Following the recommendations of \citet{Scranton2005}, for galaxy magnitudes we use the SDSS \texttt{cModelMag} magnitudes, and scale magnitude errors according to Eq. 15 in \citet{Scranton2005}, while for galaxy colours we use SDSS \texttt{modelMag} magnitudes. Similarly to other SDSS applications, we adopt WMAP 5-year + SNe + BAO best-fitting cosmological parameters: $\Omega_\Lambda = 0.726$, $\Omega_m = 0.2739$, $\Omega_r = 0.0001$ and $H_0=70.5 km/s/Mpc$ \citep{Hinshaw2009}.

The photometric database can be accessed via \textit{SkyServer}\footnote{\url{http://skyserver.sdss.org/CasJobs/}}. More information on the data used for this study, and program source code are available on the web site of the paper\footnote{\url{http://www.vo.elte.hu/papers/2016/photoz/}}. Colour versions of the figures are available in the online version of the paper.

\subsection{Photometric redshift estimation}

In the literature, there are two main approaches to estimating redshifts from broad-band photometry: the empirical, and the template-based approach. 

Empirical methods generally utilise a supervised machine learning algorithm to find patterns in a training set -- with both broad-band magnitude and redshift values -- that allow prediction for cases when the redshift is not known. The 'similarity' of galaxies is usually defined in a metric space, the dimensions of which are some combination of the broad-band magnitudes and colours, perhaps with some scaling applied -- we will refer to this as the colour and magnitude space, or, even more concisely, the colour space. The metric is generally chosen to be Euclidean distance within the colour space. Galaxies with a small distance between them -- i.e. local galaxies -- are considered to be similar, therefore their redshifts are also assumed to be similar. This assumption is then used in an algorithm for estimating galaxies with unknown redshifts. Examples of machine learning tools used for this purpose include artificial neural networks \citep{Collister2007, Reis2012, Brescia2014}, local polynomial fits \citep{Csabai2007}, random forests \citep{Carliles2010}, and boosted decision trees \citep{Gerdes2010}.

The template-based approach generally starts with a set of spectral templates and filter transmission curves, computes synthetic photometric magnitudes from them at various redshifts, and records the redshifts of templates that best reproduce the observed photometry. The choice of spectral templates is a crucial element of these methods. Galaxy templates can be computed theoretically using stellar models, an assumed initial mass function and stellar evolutionary tracks, which is a process known as stellar population synthesis \citep{Pegase1, BruzualCharlot2003, Maraston2011, Vazdekis2012}. The modelling of emission lines in such models -- which can contribute significantly to broad-band magnitudes \citep{Atek2011} -- is a difficulty because of the number of extra parameters needed to model the interstellar medium, but additional theoretical assumptions \citep{Stasinska1984, Pegase1, cloudy} or empirical line estimation \citep{Gyory2011, Beck2016} can still be used. Alternatively, sets of measured galaxy spectra can be used to compile a library of empirical spectral templates, where the inclusion of measured lines is relatively straightforward \citep{Yip2004, Dobos2012, Marchetti2013}.

Template-based photometric redshift estimation methods in the literature include simple $\chi^2$-minimisation with a well-calibrated and wide set of templates \citep{Arnouts2002, Ilbert2006}, full Bayesian analysis using an empirical prior but relatively fewer templates \citep{Benitez2000, Coe2006}, and Bayesian analysis using a linear combination of templates \citep{Brammer2008}. Additional refinements include template corrections based on objects with known redshifts \citep{Budavari2000, Csabai2000, Feldmann2006}, and wavelength-dependent weighting of template errors \citep{Brammer2008}.

The template-based approach has notable advantages over the empirical one: a training set with known redshifts is not required, and additional physical properties are implicitly estimated, since the entire template spectral energy distribution (SED) is known. However, unknown systematics in the photometric measurements are not accounted for, as opposed to empirical methods, where these are contained within the training set. Additionally, empirical techniques generally perform considerably better than template-based ones within the object type and redshift coverage of the training set \citep{Csabai2003}. However, the extrapolating capabilities of empirical methods are typically poor.

As in previous releases, to utilise the extensive spectroscopic sample of the SDSS, we elected to use an empirical method for estimating the redshift and its error, local linear regression. To get the best of both worlds, we combined this with a template fitting step that uses the photometric redshift, yielding additional physical information. We detail our methods in Sec.~\ref{sec:methods}.

\subsection{Difficulties in photometric redshift estimation}
\label{sec:errorintro}

There are two main factors that are detrimental to the accuracy of photometric redshift estimation, regardless of the specific approach taken: the overlap in photometric colour space between different galaxy types, and the measurement errors in the photometry. While these are of different origin, their effect is very much intertwined.

The first factor, overlap in broad-band colour space, is a purely physical phenomenon. When the available colours cannot differentiate between morphological types, i.e. when different galaxy types have the same colours at different redshifts, there simply is not enough data to give an unequivocal answer to the question of what the redshift is. In such cases, the assumption that the broad-band magnitudes and colours uniquely determine the redshift does not hold, there are degeneracies in the colour--redshift relation \citep{Benitez2000}.

The second factor, photometric measurement errors, is a major issue. The measurement errors can greatly exacerbate the effects of overlap, blurring the divisions in colour space between different galaxy types, and also between galaxies of the same type but with differing redshifts \citep{Benitez2000}. Additionally, when the measurement errors are not estimated accurately, or when photometric errors in different bands are correlated \citep{Scranton2005}, the assumption of uncorrelated Gaussian errors, used in many methodologies, simply does not hold \citep{Budavari2009}.

These issues can be mitigated by improvements in the instrumentation. A better camera and telescope can reduce photometric errors \citep{Ivezic2008, Tonry2012}, while a large selection of filters \citep{Wolf2003}, or filters designed specifically for photometric redshift estimation \citep{Budavari2001} can remove degeneracies.

In Sec.~\ref{sec:errordiscussion}, we discuss how these factors affect our results.

\section{Methods}
\label{sec:methods}

\subsection{Local linear regression}
\label{sec:linreg}

Following \citet{Csabai2007} and earlier SDSS releases, we adopted a local (or piecewise) linear model to describe how the redshifts of galaxies depend on broad-band colours and magnitudes. The locality allows the model to follow the complex relationship between these properties, while using a polynomial of just the first order means that a relatively small number of galaxies is enough to fit the parameters. Taking just a few neighbouring galaxies into account helps preserve the local aspect of the model, and, as opposed to a simple average of the neighbours, the linear fit can follow subtle colour-dependent trends in the redshift.

Let $i$ be the index of a galaxy in the set $Q$ of galaxies to be estimated (query set), and let us denote the redshift of the $i$-th galaxy with $z_i$ and its coordinates in the $D$-dimensional colour and magnitude space with the vector $\vecvar{d}_i$. Let us use $j$ to index galaxies in the training set $T$, which is a collection of galaxies with both coordinate and redshift measurements -- $\vecvar{d}_j$ and $z_j$, respectively. Thus, our local linear model can be formulated in the following way:

\begin{equation}
\label{eq:redshift}
z_{i} \approx c_i + \vecvar{a}_i \vecvar{d}_i = z_{\mathrm{phot},i}
\end{equation}

$z_{\mathrm{phot},i}$ denotes the photometric redshift estimate. The parameter $c_i$ is a constant offset, while components of the vector $\vecvar{a}_i$ are linear coefficients. These parameters describe our model in the local neighbourhood of galaxy $i$ -- to determine them, we need to extract the local empirical relationship present in the training set, $T$. We do this by first finding the $k$-nearest neighbours of galaxy $i$ within $T$, i.e. the $k$ galaxies whose $\vecvar{d}_j$ coordinates are the closest to $\vecvar{d}_i$ in terms of Euclidean distance. Let us denote the set of nearest neighbours by $\nn$. The parameters can then be determined using standard linear regression, by minimising the expression

\begin{equation}
\chi^2_i = \sum_{j \in \nn} \frac{\left(z_j - c_i - \vecvar{a}_i \vecvar{d}_j \right)^2}{w_j}
\end{equation}

where $w_j$ is a weight that could e.g. represent uncertainties in $z_j$ and $\vecvar{d}_j$, or it could be a function of the distance between $\vecvar{d}_i$ and $\vecvar{d}_j$. The summation runs over the nearest neighbours, and the $\chi^2_i$-minimisation has to be done for every galaxy $i$ within $Q$. The error of the photometric redshift $z_{\mathrm{phot},i}$ can be estimated by how well the thus fitted hyperplane reproduces the $z_j$ redshifts of the nearest neighbours -- we compute the RMS of the deviations from the fit:

\begin{equation}
\label{eq:redshifterror}
\delta z_{\mathrm{phot},i} \approx \sqrt{\frac{\sum_{j \in \nn} \left(z_j - c_i - \vecvar{a}_i \vecvar{d}_j \right)^2}{k}}
\end{equation}

In our current implementation, we have $D=5$ dimensions, and the components of the vectors $\vecvar{d}_i$ and $\vecvar{d}_j$ are the $r$-band magnitude, and the $u-g$, $g-r$, $r-i$, $i-z$ colours. All five dimensions are scaled to have zero mean and unit standard deviation. The nearest neighbours are weighted equally, $w_j=1$ for every $j$. These choices were made to optimise the accuracy of the photo-z estimation. We use $k=100$ to have enough data points to determine the parameters and the error, but still preserve the locality of the model. The exact choice of $k$ does not significantly impact the results, however.

We assume that the error of the spectroscopic redshift is negligible, i.e. $z_j=z_{\mathrm{spec},j}$. \revision{Generally, this} is a reasonable approximation because spectroscopic redshifts are much more accurate than photometric redshift estimates. \revision{However, it is important to note that there is a non-negligible percentage of spectroscopic redshift failures corresponding to a given quality cut in a survey (see Sec.~\ref{sec:crossmatch} for a discussion of failure rates). If the failures are correlated with spectral type and colour, this systematic error in $z_{\mathrm{spec}}$ will be included in our training set, and thus propagate through to our $z_{\mathrm{phot}}$ estimates. Still, our best reference points for estimation are the redshifts published by spectroscopic surveys.}

As an additional refinement of our method, when there are neighbours with outlying redshifts, we perform the computations twice to eliminate them. Neighbours that satisfy $3\delta z_{\mathrm{phot},i}<|z_j - c_i - \vecvar{a}_i \vecvar{d}_j|$ are discarded from the set $\nn$, and the fit is redone for the limited set of $l<k$ nearest neighbours, as needed. Also, we flag galaxies that lie outside the bounding box of the nearest neighbours in the $D$-dimensional colour and magnitude space. In such cases, we perform an extrapolation using the fitted hyperplane as opposed to an interpolation, therefore we can expect less reliable results (see Sec.~\ref{sec:errordiscussion} for more details).

Once the photometric redshift of the query point has been determined using this empirical method, we follow up with a spectral template fitting step, as described in the following section.

\subsection{Spectral template fit}
\label{sec:templatefit}

Let us denote redshift with $z$, galaxy type with $t$, measured magnitudes and magnitude errors with $m$ and $\Delta m$, and synthetic magnitudes with $s$. Let us index the $D$-dimensional magnitude space with $p$, and, again, index galaxies in the query set with $i$.  Under this notation, traditional maximum likelihood template-based photometric redshift estimation methods \citep{Bolzonella2000, Csabai2000, Arnouts2002, Ilbert2006} solve the following problem:

\begin{equation}
\left(z_{\mathrm{phot},i}; t_i; m_{0,i}\right) = \argmin_{\left(z; t; m_0\right)}\sum_{p = 1}^{D} \left(\frac{ m_{p,i} - \left(s_p\left(z,t\right) - m_0\right)}{\Delta m_{p,i}}\right)^2
\end{equation}

Here the constant offset in magnitude, $m_0$, is a scaling factor for the total flux with respect to the synthetic total flux. Generally, $z$ and $t$ iterate over a pre-determined list of redshifts and galaxy templates, while the best-fitting $m_0$ can be calculated analytically for a given $z$ and $t$.

\begin{figure}
	\includegraphics{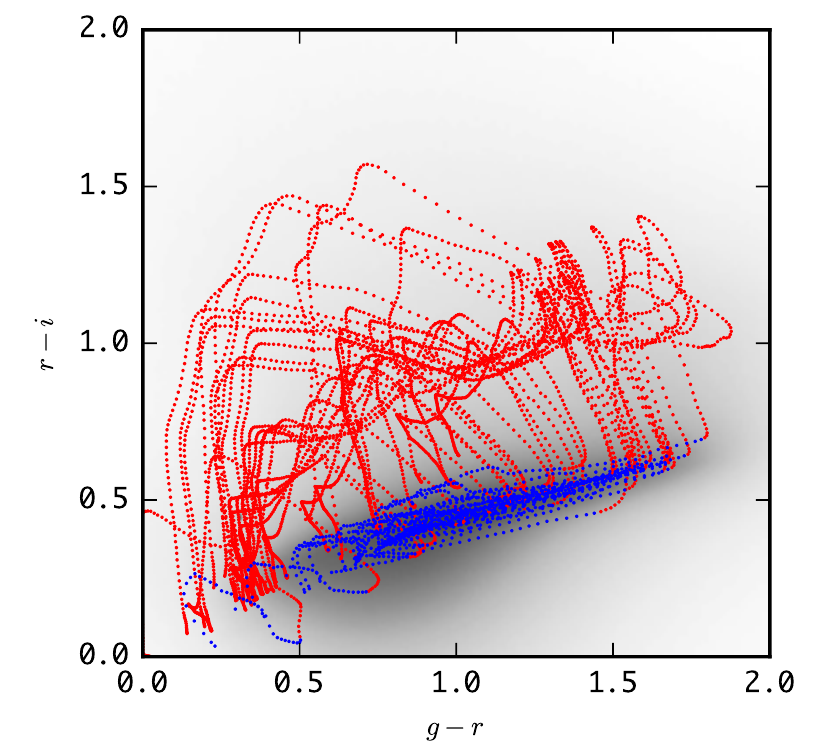}
	\caption{\revision{The colour space coverage of the spectral templates from \citet{Dobos2012} in $g-r$, $r-i$ dimensions. Blue dots show templates with $z<=0.35$, while red dots correspond to redshifts $0.35<z<0.7$. The template colours are superimposed on a grayscale density map of SDSS photometric measurements.}}
	\label{fig:tracks}
\end{figure}

In our hybrid approach, instead of iterating over $z$, we use the empirically determined photometric redshift, as described in Sec.~\ref{sec:linreg}. This way, we enjoy the benefit of higher redshift accuracy due to the extensive training set, while also fitting a galaxy template with a known SED. Thus, the expression we solve becomes:

\begin{equation}
\label{eq:templatefit}
\left(t_i; m_{0,i}\right) = \argmin_{\left(t; m_0\right)}\sum_{p = 1}^{D} \left(\frac{ m_{p,i} - \left(s_p\left(z=z_i,t\right) - m_0\right)}{\Delta m_{p,i}}\right)^2
\end{equation}

where $z_i$ is computed using Eq.~\ref{eq:redshift}. As for the list of templates, we use the composite spectrum atlas of \citet{Dobos2012}, which has been assembled from SDSS spectra, takes emission lines into account, and contains extreme red and blue galaxy types in addition to the more frequently occuring ones. \citet{Dobos2012} also published synthetic photometric magnitudes in the SDSS filter set for a grid of redshift values. \revision{Fig.~\ref{fig:tracks} shows the coverage of the templates in $g-r$, $r-i$ colours -- the dense galaxy regions are well-covered by the composite spectrum atlas. For the set of all photometric galaxies, the fitted synthetic magnitudes are within $3 \,\Delta m$ of the measured $m$ for $82.0\%$, $89.7\%$, $96.2\%$, $97.2\%$ and $97.3\%$ of cases, respectively, for the $u$, $g$, $r$, $i$ and $z$ broad-band magnitudes. The normalized error distributions are roughly Gaussian, with the exception of the $u$-band, where it is asymmetric. Considering the redshift estimation errors and outlier rate in the unfiltered galaxy set (see Sec.~\ref{sec:results} and Tab.~\ref{tab:errorclasses2} for more details), the within-$3 \,\Delta m$ ratios are relatively high, which shows that the templates adequately describe the fitted galaxies.}

Once we have found the best-fitting spectral template $t_i$, we determine other values of physical interest. The $DM$ distance modulus and $D_L$ luminosity distance are computed using the redshift and our assumed cosmology. Knowing the SED of the template, we also calculate observed-frame synthetic magnitudes, K-corrections to redshifts $0$ and $0.1$, and rest-frame absolute magnitudes (see Sec.~\ref{sec:appendix1} for exact definitions of these).


\section{Training set}
\label{sec:trainingSet}

Our training set initially consisted of the entire spectroscopic galaxy catalog of the Sloan Digital Sky Survey Data Release 12. This includes the earlier main galaxy and LRG samples, and also the more recent BOSS sample. The main galaxy sample consists of a wide variety of galaxies, with no cuts on colour, although it is rather limited in terms of redshift \citep{Strauss2002}. The LRG sample provides an expanded redshift coverage, however, it has specifically targeted luminous red galaxies \citep{Eisenstein2001}. The BOSS sample extends much deeper than the former two, and has somewhat relaxed the sharp colour cuts of the LRG sample, but it is still targeted towards massive galaxies \citep{Dawson2013}, likely resulting in non-negligible selection effects.

On the other hand, the photometric galaxy catalog of the SDSS has no such selection effects, and our goal is to provide photometric redshifts for the entire catalog, not just a subset of morphological types or colour. Since it would be advantageous to have a wider selection of galaxy types and colours even at higher redshifts, we decided to extend our training set by cross-matching galaxies in the Sloan photometric catalog with spectroscopic measurements of other, publicly available surveys.

\subsection{Data from other surveys}
\label{sec:crossmatch}

The spectroscopic surveys with which we extended our training set are listed in Tab.~\ref{tab:surveys}, with references. For each survey, we used the published redshift quality flag to select only the reasonably confident redshift measurements, with confidences $\geq 95\%$ (with the exception of PRIMUS, where $\geq 92\%$).

\begin{table*}
	\begin{tabular}{c | c | c }
		\hline
		
		Survey Name & References & Quality flag \\
		
		\hline
		
		2dF & \citet{Colless2001, Colless2003} & $Quality=4,5$ \\
		
		6dF & \citet{Jones2004, Jones2009} & $Q=3,4$ \\
		
		DEEP2 & \citet{Davis2003, Newman2013} & $ZQUALITY=3,4$ \\
		
		GAMA & \citet{Driver2011, Baldry2014} & $NQ=4$ \\

		PRIMUS & \citet{Coil2011, Cool2013} & $Q=4$ \\	

		VIPERS & \citet{Garilli2014, Guzzo2014} & $\left \lceil{zflg}\right \rceil \bmod 10=3,4$ \\
		
		VVDS & \citet{LeFevre2004, Garilli2008} & $ZFLAGS \bmod 10=3,4$ \\
		
		WiggleZ & \citet{Drinkwater2010, Parkinson2012} & $Qop=4,5$ \\
		
		zCOSMOS & \citet{Lilly2007, Lilly2009} & $\left \lceil{Class}\right \rceil \bmod 10=3,4$ \\		
		
		\hline
	\end{tabular}
	\caption{Information about the external spectroscopic surveys we used to expand our training set.}
	\label{tab:surveys}
\end{table*}

We cross-matched the galaxies from other surveys with SDSS primary photometric galaxy measurements by using J2000 right ascension and declination coordinates and published astrometric errors. We followed the probabilistic methodology of \citet{Budavari2008}, assumed Gaussian errors, and calculated the Bayes factor of Eq.~16 in \citet{Budavari2008}, which is the ratio of the likelihood that the two measurements are of the same source, and the likelihood that they are of separate sources:

\begin{equation}
	B = \frac{L\left(\textrm{same source}\right)}{L\left(\textrm{separate sources}\right)}= \frac{2}{\sigma_1^2 + \sigma_2^2}exp\left\{-\frac{\psi^2}{2\left(\sigma_1^2 + \sigma_2^2\right)}\right\}
	\label{eq:crossmatchBayes}
\end{equation}

Here $\sigma_1$ and $\sigma_2$ are the astrometric errors of two given galaxies, and $\psi$ is the angular separation between them. We accepted matches with $B>10,000$, thus ensuring that we only used rather certain matches.

Galaxies with existing SDSS spectrometry were excluded from the cross-match, and where we found multiple matches for the same Sloan galaxy, we selected the one with the smallest redshift error.

In total, we found $168,834$ matches with reasonable redshift confidence. However, later filtering steps greatly limited the number we could utilise, to $76,193$.

\revision{Multiple matches provide an opportunity to test whether the published spectroscopic redshift failure rates are correct and whether the cross-match itself works reliably. Of the total of $1,012$ multiple matches in the filtered sample, $171$ were between two PRIMUS measurements, $769$ had one PRIMUS object, and $72$ did not include PRIMUS. (We are handling PRIMUS separately because of its lower confidence level and higher redshift error compared to other surveys.)}

\revision{The only-PRIMUS set had a standard deviation of $\sigma\left(\Delta z_{\mathrm{spec}}\right) = 0.00473$ and a $3\sigma$ outlier rate of $P_o = 9.36 \%$ (outliers were removed iteratively). Individual PRIMUS measurements typically have an accuracy of $\sim 0.005$, therefore the deviation is even below what one would expect, and the outlier rate is also well below the theoretically expected $1 - 0.92 \times 0.92 = 15.4\%$.}

\revision{The PRIMUS--other survey set is described by the numbers $\sigma\left(\Delta z_{\mathrm{spec}}\right) = 0.00472$ and $P_o = 8.97 \%$. The deviation is roughly the accuracy of PRIMUS, just as expected, while the outlier rate is again below the expected $1 - 0.95 \times 0.92 = 12.6\%$.}

\revision{The non-PRIMUS set had a standard deviation of $\sigma\left(\Delta z_{\mathrm{spec}}\right) = 0.00071$ and an outlier rate of $P_o = 29.2 \%$. The deviation is negligible for our purposes, but the outlier rate is significantly larger than the expected $1 - 0.95 \times 0.95 = 9.75\%$. The observed discrepancies might be due to a number of reasons, below we list a few.}

\revision{
\begin{itemize}
	\item The spectroscopic redshift failures could be correlated, which would reduce the combined outlier rate. Especially the only-PRIMUS set could be affected, where the same survey measured the same object twice.
	\item Galaxies could be erroneously cross-matched due to an underestimation of astrometric accuracy -- DEEP2, the survey responsible for $71.4 \%$ of outliers in the non-PRIMUS set, uses the Canada France Hawaii Telescope, which quotes the USNOA 2.0 astrometric error of $0.5^{\prime\prime}$ \citep{Coil2004}, the highest value of all the external surveys.
	\item Overlapping galaxies in the field of view could compromise spectroscopic redshifts, and could also lead to incorrect cross-matches.
\end{itemize}}

\revision{Furthermore, it is important to note that the non-PRIMUS set only had $72$ matches, of which $21$ were outliers. This is a rather small sample size, and might not be representative. On the whole, the confidence levels derived from multiple matches are in line with -- or better than -- the expectations based on the published numbers of the surveys.}

\subsection{Filtering the training set}
\label{sec:filtering}

While our goal was to assemble a training set with as wide a coverage in redshift and colour space as possible, the inclusion of objects with too large photometric errors would diminish our ability to find the most similar reference galaxies. The 'true' nearest neighbours may be scattered away due to errors, with less similar galaxies taking their place. To alleviate this problem, we introduced photometric error cuts to the training set. Additionally, we filtered out galaxies with outlying colours, which both eliminates erroneous measurements, and also more clearly defines the boundaries of our training set in colour space.

The exact parameters of the cuts were determined empirically, with the following criteria in mind:

\begin{itemize}
	\item optimise the photometric redshift estimation results,
	\item leave no region empty in the space of broad-band colours, if otherwise within the coverage of the training set,
	\item keep fainter and higher-redshift measurements of sufficient accuracy.
\end{itemize}

The final values of the photometric error and colour cuts are as follows:

\begin{equation}
\begin{aligned}
	\Delta r &< 0.15 \\
	\Delta (g-r) &< 0.225 \\
	\Delta (r-i) &< 0.15 \\
	\Delta (i-z) &< 0.25 \\
	-0.911 < (u-g) &< 5.597 \\
	0.167 < (g-r) &< 2.483 \\
	0.029 < (r-i) &< 1.369 \\
	-0.452 < (i-z) &< 0.790 \\
\label{eq:cuts}	
\end{aligned}
\end{equation}

Magnitudes are in the SDSS \texttt{ugriz} filter system, with errors scaled following \citet{Scranton2005} (see also Sec.~\ref{sec:intro}). The colour cuts correspond to filtering out the highest and lowest $0.5\%$ of data for the $(u-g)$ colour, and $1\%$ for the other three colours. The reason for having no limit on $\Delta u$, and for having relaxed criteria on $(u-g)$ compared to other colours is that the errors of the SDSS $u$-band are generally much larger than that of other bands, and even galaxies with fairly secure photometric redshifts can have very large $u$-band errors.

\begin{figure}
	\includegraphics{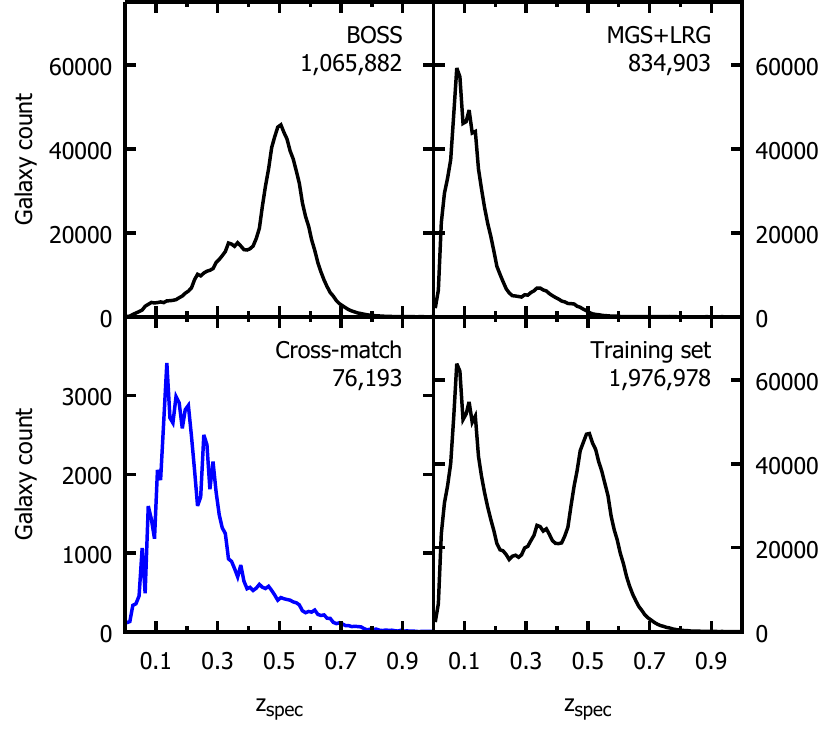}
	\caption{The redshift distribution of our entire training set, and subsets of it: the BOSS spectroscopic sample, the pre-BOSS spectroscopic sample that includes the main galaxy sample and the LRG sample (MGS+LRG), and the additional galaxies cross-matched from other surveys. In the top right corner of each panel, we indicate the corresponding subset, and the total number of galaxies within that subset. Note the different scale of the cross-match subset.}
	\label{fig:zhist}
\end{figure}

Only those galaxies were included in the training set that fulfilled all of Eq.~\ref{eq:cuts}. Additionally, SDSS galaxies with unsecure spectroscopic redshifts were also cut: the spectroscopic error flag \texttt{SpeczWarning} had to either take the value \texttt{OK} or \texttt{MANY\_OUTLIERS} (the latter rarely signifying a real error according to the documentation). \revision{This spectroscopic error flag cut filters out a higher and higher fraction of galaxies as the redshift increases -- with more distant galaxies typically having lower signal-to-noise spectra -- but there is no indication of a specific redshift being preferentially eliminated, which otherwise could have pointed to a systematic incompleteness in our training set.} The redshift distribution of the finalised training set is shown in Fig.~\ref{fig:zhist}.

\section{Results}
\label{sec:results}

\subsection{Accuracy of photometric redshifts}
\label{sec:accuracy}

\begin{figure*}
\begin{minipage}[t]{\textwidth}
	\includegraphics{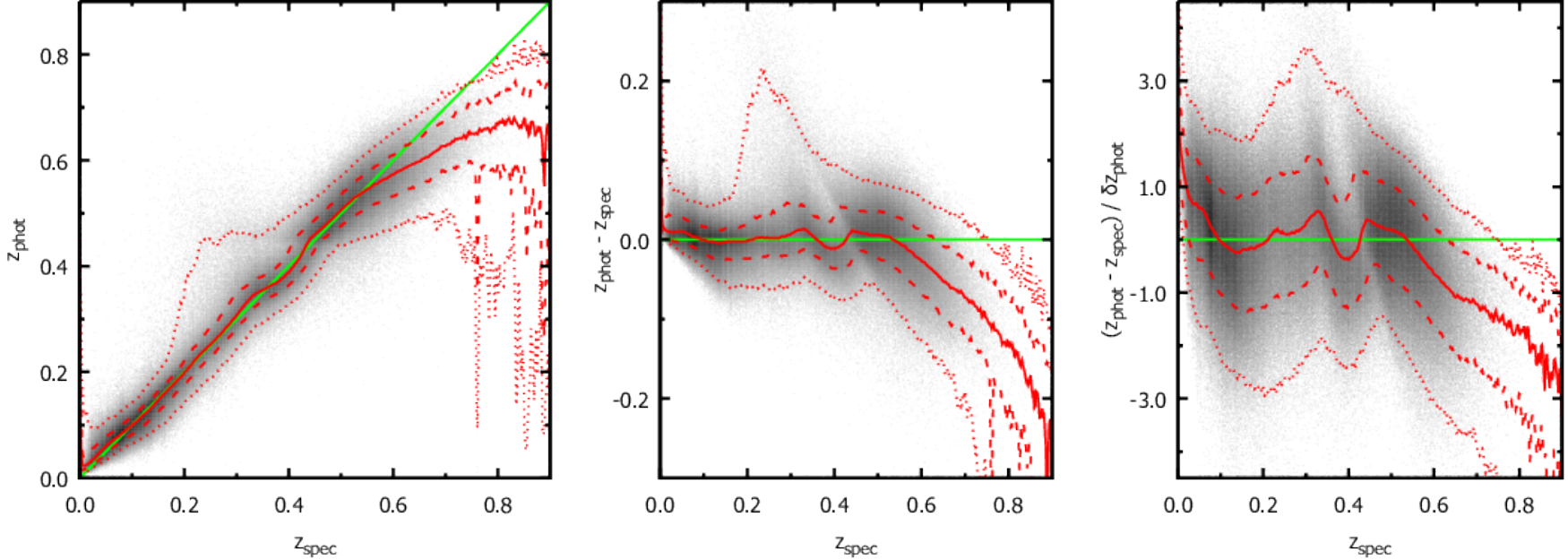}
	\caption{\revision{The photometric redshift} ($z_{\mathrm{phot}}$) as a function of spectroscopic redshift ($z_{\mathrm{spec}}$), $z_{\mathrm{phot}}-z_{\mathrm{spec}}$ as a function of $z_{\mathrm{spec}}$, and $(z_{\mathrm{phot}}-z_{\mathrm{spec}})/\delta z_{\mathrm{phot}}$ as a function of $z_{\mathrm{spec}}$, where $\delta z_{\mathrm{phot}}$ is the photometric redshift error estimate. The galaxy density of our training set is shown in grayscale -- we took the logarithm of galaxy counts so that even individual galaxies can be seen. The red solid, dashed and dotted lines represent the median, $68\%$ and $95\%$ confidence regions of the training set, respectively. The green line shows $z_{\mathrm{spec}}=z_{\mathrm{phot}}$, i.e. what would be the perfect estimation. See the text for a discussion.}
	\label{fig:zscatter}
\end{minipage}
\end{figure*}

To evaluate the performance of our methods, we randomly divided the training set into two equal-sized subsets, and performed cross-validation, estimating the photometric redshifts of one half using the other half as the training set (and vice versa). The resulting photometric redshifts could then be contrasted with the spectroscopic redshifts. \revision{Fig.~\ref{fig:zscatter}} shows the photometric redshift $z_{\mathrm{phot}}$, the estimation error $z_{\mathrm{phot}}-z_{\mathrm{spec}}$, and the estimation error divided by the reported photometric redshift error $\delta z_{\mathrm{phot}}$, as functions of the spectroscopic redshift.

Using the normalized redshift estimation error $\Delta z_{\mathrm{norm}} =  \frac{z_{\mathrm{phot}} - z_{\mathrm{spec}}}{1+z_{\mathrm{spec}}}$, we achieve an average bias of $\overline{\Delta z_{\mathrm{norm}}} = \revision{5.84 \times 10^{-5}}$, a standard deviation of $\sigma\left(\Delta z_{\mathrm{norm}}\right)=\revision{0.0205}$, and an outlier rate of $P_o=\revision{4.11\%}$. Outliers are defined as $|\Delta z_{\mathrm{norm}}| > 3\sigma\left(\Delta z_{\mathrm{norm}}\right)$, and are removed iteratively. While most galaxies in the training set have fairly small estimation errors, on \revision{Fig.~\ref{fig:zscatter}} it is apparent that there are redshift ranges where there is a non-negligible bias, up to $\Delta z=0.01$ or $0.5 \, \delta z_{\mathrm{phot}}$.

Still, in biased regions \revision{between $58\%$ and $76\%$ of galaxies are within $\pm \, 1 \, \delta z_{\mathrm{phot}}$, and between $86\%$ and $98\%$ of galaxies are within $\pm \, 2 \, \delta z_{\mathrm{phot}}$}. Thus, the confidence interval $z_{\mathrm{phot}} \pm \delta z_{\mathrm{phot}}$ can be \revision{reasonably} used in applications, as it will contain \revision{a fairly high fraction of galaxies} even when there is bias in the estimation, and the distribution is not centered on $z_{\mathrm{phot}}$.

\begin{figure}
	\includegraphics{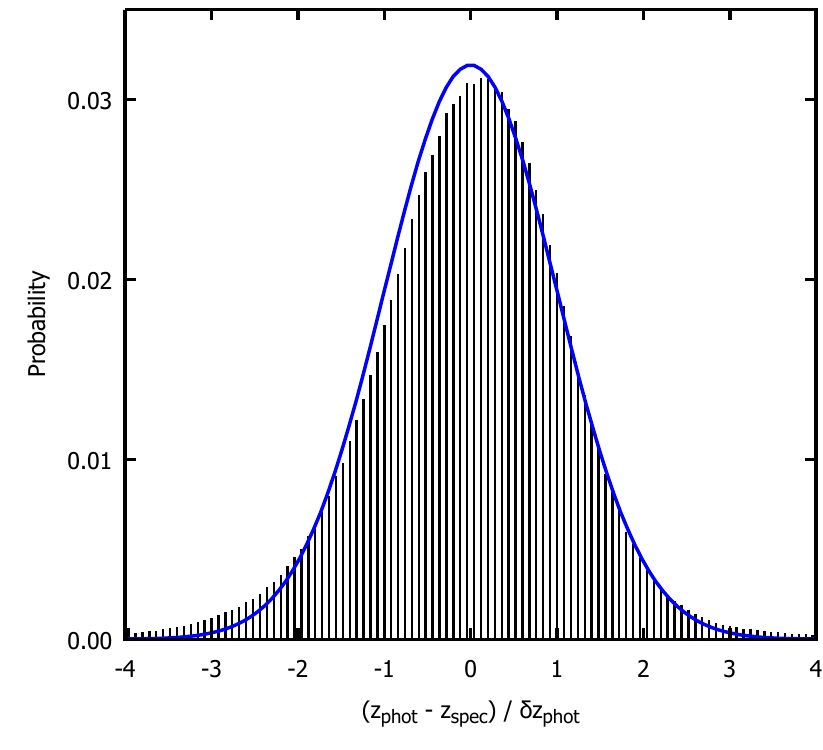}
	\caption{The normalized histogram of the scaled redshift error, $(z_{\mathrm{phot}}-z_{\mathrm{spec}})/\delta z_{\mathrm{phot}}$, is plotted in black for our training set. The blue line shows the standard normal distribution. See the text for a discussion.}
	\label{fig:scalederrorhistogram}
\end{figure}

Additionally, on Fig.~\ref{fig:scalederrorhistogram}, we plotted the probability density function of $(z_{\mathrm{phot}}-z_{\mathrm{spec}})/\delta z_{\mathrm{phot}}$ alongside a standard normal distribution. There is a small overall bias, but otherwise the two distributions match rather well, highlighting that our method for estimating the error of the photometric redshift gives a fair assessment of the estimation accuracy.

\revision{Another issue visible on Fig.~\ref{fig:zscatter} is that the estimation accuracy declines dramatically from around $z=0.6$, where the number count of the training set falls off. These high-redshift galaxies occupy sparsely sampled regions in colour space, as evidenced by the fact that $94\%$ of them are above the 50th, and $68\%$ are above the 75th percentile of nearest neighbour bounding box volume. Sparse regions are more likely to include a non-negligible amount of galaxies scattered there due to high photometric errors. In the case of high-redshift galaxy regions, the scattered galaxies are also more likely to have lower redshifts, hence the negative estimation bias.}

In the following section, we will go into more detail concerning the biases and errors.

\subsection{Discussion of biases and errors}
\label{sec:errordiscussion}

Here we discuss how the issues outlined in Sec.~\ref{sec:errorintro}, namely overlap and errors in the photometry, affect our results.

When galaxies of different redshifts overlap in colour space, the nearest neighbours that we find with our algorithm will have a bimodal, or even multimodal redshift distribution. In this case, the estimated redshift will lie between the different peaks in the distribution, and the estimated error will increase accordingly. Additionally, galaxies belonging to a peak at a smaller redshift will have a positive estimation bias, while galaxies that correspond to a peak at a higher $z$ will be estimated with a negative bias. When the overlap is 'perfect', it is impossible to decide which is the appropriate galaxy group, but when the loci of groups in colour space have a slight offset between them, the closer galaxy group will more strongly constrain the fitted hyperplane locally. Assuming a mixture of two Gaussian distributions of equal weight but with different means, our method will estimate the redshift closer to the correct peak, as opposed to a simple $k$-nearest neighbour average, which would give the centerpoint, the average of the means. This effect becomes less noticeable when one of the groups is underrepresented in the training set, or when photometric errors are large enough to sufficiently mix the groups in colour space. To remove some of the degeneracy, in addition to the colours, we also used the $r$-band magnitude in our local linear regression. In Sec.~\ref{sec:errormap}, we describe an error map that helps quantify the effects of overlap.

Photometric measurement errors strongly limit the accuracy of photometric redshifts in the SDSS catalog -- both the training set and the query set are affected, especially the fainter galaxies. We introduced several photometric error classes for the galaxies to quantify the dependence of redshift estimation errors on errors in the photometry: class $1$ matches the error limits of the training set, and the subsequent classes ($2-7$) contain galaxies with progressively higher errors. The exact limits were determined empirically based on the photometric error and redshift error distributions, with the aim of giving a sequence from useful photometric redshifts to highly inaccurate ones. The class identifiers have a negative sign if the local linear regression was an extrapolation, i.e. the estimated galaxy lay outside the bounding box of the nearest neighbours. For example, class $-1$ denotes galaxies that match the error limits of the training set, but were estimated with an extrapolation. This way, class $-1$ also includes galaxies that did not satisfy the colour cuts of Eq.~\ref{eq:cuts}, and therefore are not within the training set. Classes $2-7$ and $(-2)-(-7)$ do contain galaxies with spectroscopic redshifts, specifically those that did not fulfill the error limits of Eq.~\ref{eq:cuts} -- these galaxies can be used to test the redshift estimation accuracy in a given class.

Tab.~\ref{tab:errorclasses1} gives the photometric error limits used for each class, while Tab.~\ref{tab:errorclasses2} lists the redshift estimation bias, standard deviation, outlier rate\revision{, and the} spectroscopic \revision{and photometric} galaxy count in the classes. It is clear that higher photometric errors correspond to sharply increasing biases and deviations.

\begin{table}
	\begin{tabular}{c | c | c | c | c}
		\hline
		
		Class & $\Delta r_{max}$ & $\Delta (g-r)_{max}$ & $\Delta (r-i)_{max}$ & $\Delta (i-z)_{max}$ \\
		
		\hline
		
		1 & 0.15 & 0.225 & 0.15 & 0.25 \\
		
		2 & 0.18 & 0.25 & 0.18 & 0.28 \\
		
		3 & 0.21 & 0.30 & 0.21 & 0.31 \\
		
		4 & 0.24 & 0.35 & 0.24 & 0.34 \\
		
		5 & 0.27 & 0.40 & 0.27 & 0.37 \\
		
		6 & 0.30 & 0.45 & 0.30 & 0.40 \\
		
		\hline
	\end{tabular}
	\caption{The maximum photometric error values that a galaxy belonging to a photometric error class was allowed to have. The errors were scaled following \citet{Scranton2005}. Each galaxy is placed in the lowest possible class. Class 7 contains galaxies that could not be placed in any other class.}
	\label{tab:errorclasses1}
\end{table}

%
%
%
%
%
%
%
%
%
%

\begin{table*}
	\begin{tabular}{c | c | c | c | c | c}
		\hline
		
		Class & $\overline{\Delta z_{\mathrm{norm}}}$ & $\sigma\left(\Delta z_{\mathrm{norm}}\right)$  & $P_o$ & $N_{\mathrm{spec}}$ & \revision{$N_{\mathrm{phot}}$}\\
		
		\hline
		
		1 & \revision{$6.11 \times 10^{-5}$} & \revision{$0.0204$} & \revision{$4.07\%$} & $1,957,234$ & \revision{$42,410,836$} \\ 

		2 & \revision{$-0.0033$} & \revision{$0.0333$} & \revision{$4.03\%$} & $77,281$ & \revision{$5,657,368$}\\	
		
		3 & \revision{$-0.0057$} & \revision{$0.0331$} & \revision{$3.93\%$} & $68,610$ & \revision{$5,240,766$}\\	
		
		4 & \revision{$-0.0082$} & \revision{$0.0369$} & \revision{$4.45\%$} & $36,218$ & \revision{$3,955,814$}\\ 
		
		5 & \revision{$-0.0107$} & \revision{$0.0412$} & \revision{$5.00\%$} & $19,110$ & \revision{$2,970,417$}\\ 
		
		6 & \revision{$-0.0127$} & \revision{$0.0486$} & \revision{$5.33\%$} & $10,674$ & \revision{$2,232,881$}\\ 
		
		7 & \revision{$-0.0222$} & \revision{$0.0823$} & \revision{$3.80\%$} & $16,563$ & \revision{$6,950,249$}\\ 
		
		\hline
		
		-1 & \revision{$4.22 \times 10^{-4}$} & \revision{$0.0289$} & \revision{$5.71\%$} & $19,744$ & \revision{$2,001,544$}\\
		
		-2 & \revision{$-0.0051$} & \revision{$0.0549$} & \revision{$11.2\%$} & $5,940$ & \revision{$1,421,618$}\\
		
		-3 & \revision{$-0.0081$} & \revision{$0.0514$} & \revision{$8.97\%$} & $10,262$ & \revision{$2,848,424$}\\
		
		-4 & \revision{$-0.0104$} & \revision{$0.0567$} & \revision{$7.65\%$} & $11,200$ & \revision{$4,098,896$}\\
		
		-5 & \revision{$-0.0150$} & \revision{$0.0643$} & \revision{$6.68\%$} & $10,917$ & \revision{$5,118,595$}\\
		
		-6 & \revision{$-0.0165$} & \revision{$0.0728$} & \revision{$6.06\%$} & $10,350$ & \revision{$5,862,776$}\\
		
		-7 & \revision{$-0.0488$} & \revision{$0.1410$} & \revision{$2.30\%$} & $86,574$ & \revision{$117,703,892$}\\
		
		\hline
	\end{tabular}
	\caption{The average redshift estimation bias $\overline{\Delta z_{\mathrm{norm}}}$, standard deviation $\sigma\left(\Delta z_{\mathrm{norm}}\right)$, outlier rate $P_o$\revision{,} number of spectroscopic galaxies $N_{\mathrm{spec}}$ \revision{and number of photometric galaxies $N_{\mathrm{phot}}$} in each photometric error class, with $\Delta z_{\mathrm{norm}} =  \frac{z_{\mathrm{phot}} - z_{\mathrm{spec}}}{1+z_{\mathrm{spec}}}$. Outliers are defined to have $|\Delta z_{\mathrm{norm}}| > 3\sigma\left(\Delta z_{\mathrm{norm}}\right)$, and are removed iteratively. We indicate extrapolation in the local linear regression with a negative sign in front of the class identifier.}
	\label{tab:errorclasses2}
\end{table*}

We emphasise here that, since the training set only contains galaxies of class 1 or -1, the redshift error estimate ($\delta z_{\mathrm{phot}}$) is expected to be an accurate representation of the estimation error only when the query galaxy also belongs to class 1 or -1 (and satisfies Eq.~\ref{eq:cuts}, if class -1). As we show in Sec.~\ref{sec:errormap}, the redshift estimation errors are dependent on the position in colour space. A higher photometric error class leads to additional variance in $z_{\mathrm{phot}}$, which therefore should ideally be characterized as a function of the position in colour space. However, as shown in Tab.~\ref{tab:errorclasses2}, there are relatively few spectroscopic galaxies in the higher error classes, and we do not have a good enough coverage to allow a detailed treatment of this phenomenon. As a crude first approximation for other classes, the class-wide extra variance with respect to class 1 can be added according to Tab.~\ref{tab:errorclasses2}.

\subsection{The redshift error map}
\label{sec:errormap}
As described in the previous section, the presence of biases and higher errors in the redshift estimation is strongly dependent on the position of a given galaxy in the space of broad-band colours. To provide a tool for filtering out regions in the colour space where these issues are the most prominent, we compiled and published an error map.

The error map gives the redshift estimation results -- as computed on the training set -- for a 3D grid in $r$-band magnitude, and $g-r$, $r-i$ colours. For each bin in the grid, we report the galaxy count, the average $z_{\mathrm{spec}}$, the average $z_{\mathrm{phot}}$, the rms of $z_{\mathrm{phot}} - z_{\mathrm{spec}}$, the average $\delta z_{\mathrm{phot}}$, and the average standard deviation of the redshifts of the neighbours, $\sigma \left(z_{\nn}\right)$. With the help of this map, it is possible to flag galaxies in sparsely populated regions, or in regions with high estimation errors (which also indicate possible biases).

\begin{figure}
	\includegraphics{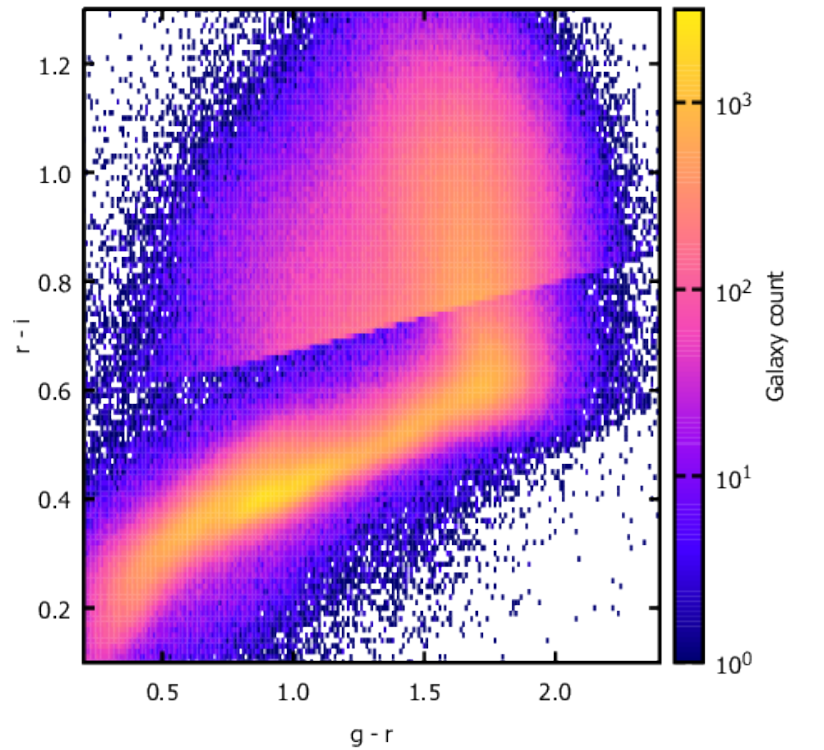}
	\caption{The galaxy count distribution of the training set, on a 2D grid of $g-r$, $r-i$ broad-band colours. The published 3D map also includes the $r$-band magnitude, this 2D version is used here for illustrative purposes. The discontinuity is caused by the colour cut of the CMASS sample \citep{Dawson2013}. See the text for a discussion.}
	\label{fig:countmap}
\end{figure}

\begin{figure*}
\begin{minipage}[t]{\textwidth}
	\includegraphics{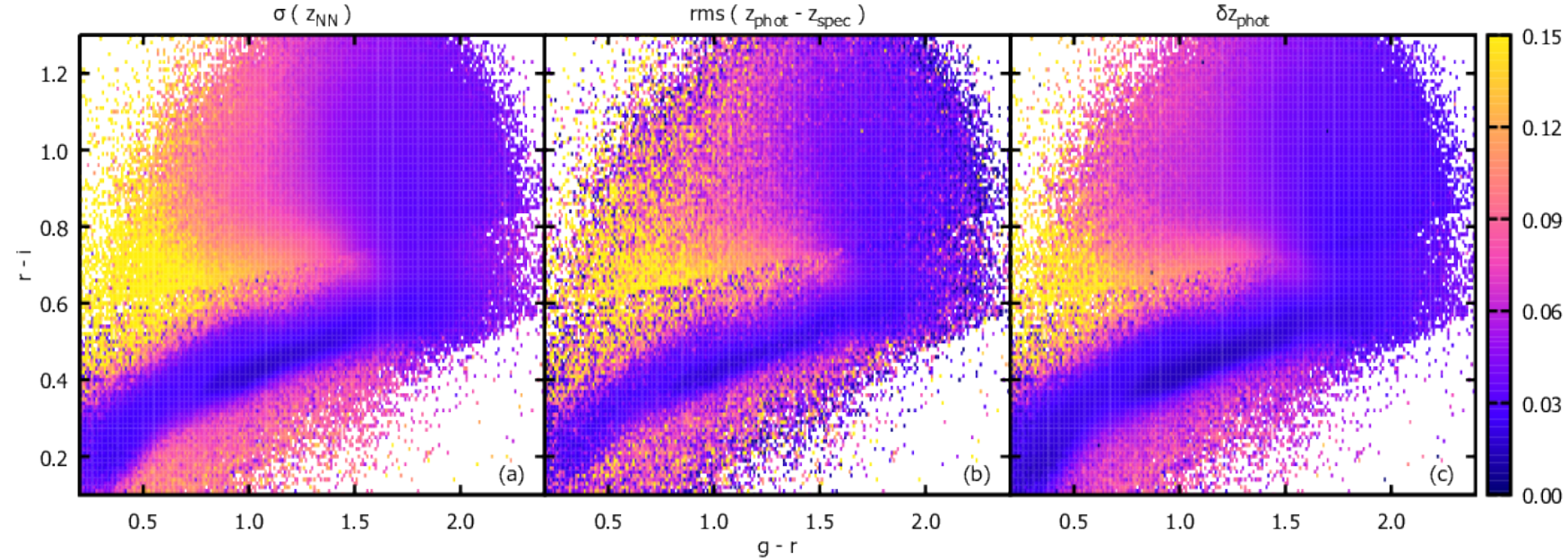}
	\caption{Photometric redshift estimation results for the training set, on a 2D grid of $g-r$, $r-i$ broad-band colours. The published 3D map also includes the $r$-band magnitude, this 2D version is used here for illustrative purposes. Panel (a) shows the average standard deviation of the redshifts of the nearest neighbours ($\sigma \left(z_{\nn}\right)$), panel (b) displays the rms of $z_{\mathrm{phot}} - z_{\mathrm{spec}}$, the actual estimation error, while panel (c) shows the average reported estimation error ($\delta z_{\mathrm{phot}}$). We note that the outliers have not been removed from the rms computation, therefore panel (b) is noisier. See the text for a discussion.}
	\label{fig:errormap}
\end{minipage}
\end{figure*}

 To illustrate, we computed these measures for a 2D projection of the 3D map, where the $r$-band magnitude has been summed over, and the grid remains in $g-r$, $r-i$ colours. In Fig.~\ref{fig:countmap}, the galaxy count distribution is shown as a function of the two colours. The pronounced discontinuity -- a diagonal line -- is a target selection effect produced by the colour cut of the CMASS subsample in the BOSS survey \citep{Dawson2013}, leading to a sparsely populated region below the cut. In Fig.~\ref{fig:errormap}, we plotted three measures of the redshift error, all of which show similar behaviour. The estimation error is highest where the redshifts of the neighbours have a larger deviation, i.e. where there is overlap of galaxies with differing redshifts. Since the photometric error limit of even error class 1 is as high as $\Delta (g-r)_{max}=0.225$ and $\Delta (r-i)_{max}=0.15$ in these two dimensions, such mixing between different redshifts is to be expected. The reported error follows the actual error closely, drawing the same overall picture of the dependence of estimation errors on the location in colour space, which also supports our result in Sec.~\ref{sec:accuracy} that our redshift error estimates are accurate. Additionally, sparsely populated regions in Fig.~\ref{fig:countmap} correspond to higher errors in Fig.~\ref{fig:errormap}. Since sparse regions could be occupied either by exotic galaxy types or galaxies that were scattered there due to high photometric errors, it is not surprising that their redshift estimation is inaccurate.

\section{Using the database}
\label{sec:database}

The photometric redshift database has been made public along with the SDSS DR12. It can be accessed via \textit{SkyServer}\footnote{\url{http://skyserver.sdss.org/CasJobs/}}, it is the Photoz table within the DR12 context. The redshift error map is contained in the table PhotozErrorMap, also in the DR12 context. Refer to Sec.~\ref{sec:appendix} for a description of each column in these tables. 

\subsection{Best practices}
\label{sec:practices}

Here we intend to give a few pointers on how best to utilise our database.

As a first step, we recommend only using galaxies with a photometric error class of 1, because that is when the redshift error estimate is expected to be accurate. If more galaxies are desired, follow the instructions at the end of Sec.~\ref{sec:errordiscussion} and use Tab.~\ref{tab:errorclasses2} for including additional error classes.

When nearest neighbours in the local linear regression are outliers, they are excluded from the fit. However, having too many outliers may indicate that the given galaxy is difficult to estimate, therefore a limit should be put on the minimum number of nearest neighbours used out of the total of $k=100$, e.g. a minimum of $l=97$. Additionally, the desired accuracy may be achieved with a cut based on the redshift error estimate, e.g. only using galaxies with $\delta z_{\mathrm{phot}}<0.03$.

The linear fitting algorithm can fail when it encounters a singular or near-singular matrix -- such cases are indicated by $z_{\mathrm{phot}}=-9999$, therefore those galaxies should be excluded. If required, the redshift of the first nearest neighbour, and the average redshift of the $100$ nearest neighbours are still available, however, in this case, there is no redshift error estimate (also flagged with $\delta z_{\mathrm{phot}}=-9999$). 

When small biases are a critical issue, the redshift error map of Sec.~\ref{sec:errormap} should be used for leaving out galaxies that are located in a high-error region in colour space, but that otherwise have a low reported $\delta z_{\mathrm{phot}}$.

Additionally, the volume of the bounding box of the nearest neighbours in the colour space could also be used for filtering -- a volume that is very large means that galaxies of very different colours are used in the local linear regression, therefore the estimated redshift could be compromised. \revision{To give an idea of potential limits, a bounding box volume cut of $\texttt{nnVol}<2$ filters out $\approx 2.5\%$ of training set galaxies that are in the sparsest colour space regions, $\texttt{nnVol}<1$ eliminates $\approx 5\%$, while $\texttt{nnVol}<0.45$ cuts $\approx 10\%$. The galaxies thus filtered out have estimation accuracies of $\sigma \left(\Delta z_{\mathrm{norm}}\right)=0.0464$, $0.0400$ and $0.0342$, respectively, with the $3\sigma$ outlier rate around $7\%$ for all three cases.}

\begin{figure*}
\begin{minipage}[t]{\textwidth}
	\includegraphics{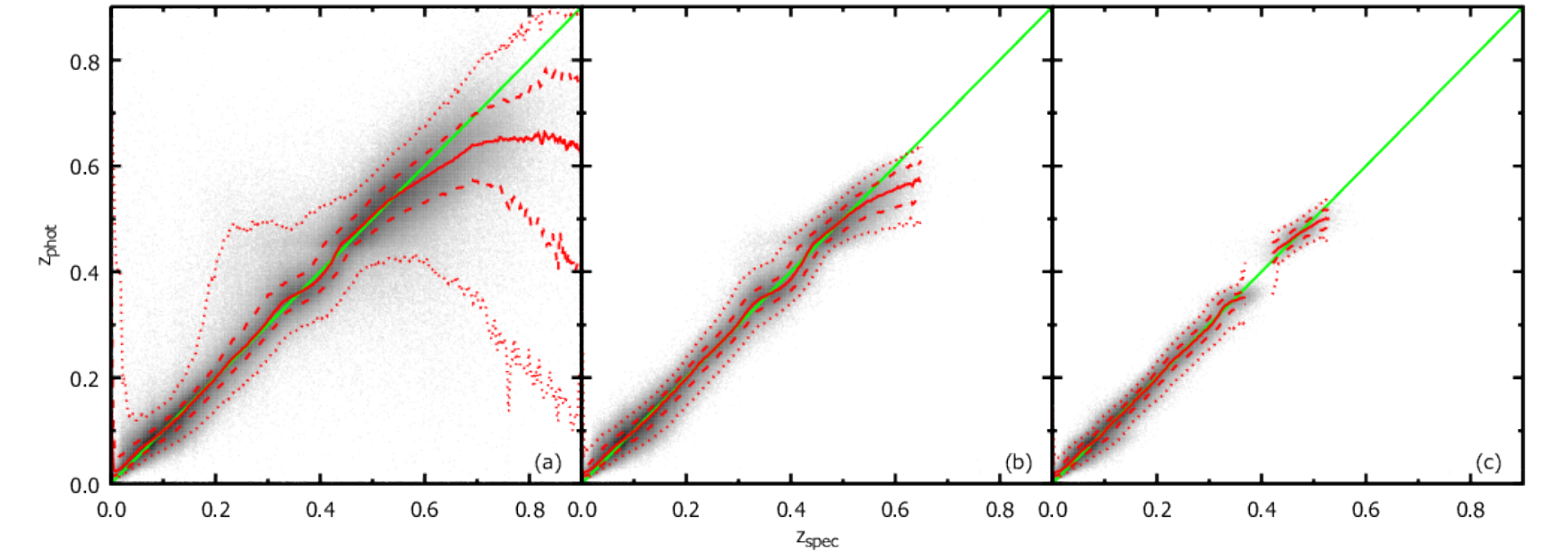}
	\caption{\revision{The photometric redshift} ($z_{\mathrm{phot}}$) as a function of spectroscopic redshift ($z_{\mathrm{spec}}$), for three different subsets of all available spectroscopic measurements. The galaxy density is shown in grayscale -- we took the logarithm of galaxy counts so that even individual galaxies can be seen. The red solid, dashed and dotted lines represent the median, $68\%$ and $95\%$ confidence regions of the data, respectively. On panel (a), there is no selection, all galaxies are shown. On panel (b), we only included galaxies of photometric error class 1 or -1, and with a reported redshift error of $\delta z_{\mathrm{phot}}<0.03$. On panel (c), galaxies of photometric error class 1 and with $\delta z_{\mathrm{phot}}<0.02$ are shown. We note that on panel (c), the more biased region around $z_{\mathrm{spec}}=0.4$ has been almost completely filtered out. See the text for a discussion.}
	\label{fig:photozcuts}
\end{minipage}
\end{figure*}

In \revision{Fig.~\ref{fig:photozcuts}}, we illustrate the hazards of using non-filtered photometric redshift data. Without implementing any cuts, the errors and the number of catastrophic failures are visibly much larger. Also, it is important to remember that we are only analysing spectroscopic measurements, which on average have much more accurate photometry than the rest of the SDSS photometric catalog -- the fraction of catastrophic outliers is expected to be much larger for the unfiltered photometric sample. On the other hand, using too stringent cuts might unneccessarily limit the redshift coverage, or the colour space coverage of the sample. For this reason, we recommend experimenting with different filtering choices to find the one most appropriate for the task at hand.

\section{Summary}
\label{sec:summary}

We described in detail how we created the photometric redshift database of SDSS DR12.

After a brief overview of the photometric redshift estimation literature, we defined the local linear regression method that we use for the redshift and redshift error estimation, and described the spectral template fitting step that followed it. We gave an account of the data and methods that went into assembling the training set. We evaluated the accuracy of our estimation via cross-validation on the training set, then we discussed the errors and biases that we encountered. We introduced photometric error classes, and a 3D redshift error map to help quantify the errors and filter out inaccurately estimated galaxies. We also provided recommendations for using the database, and choosing appropriate filtering criteria.

Our photometric redshift estimates are relatively accurate, with a standard deviation of $\sigma \left(\Delta z_{\mathrm{norm}}\right)=\revision{0.0205}$, and \revision{an acceptable} $3\sigma$ outlier rate of $P_o=\revision{4.11\%}$. The reported redshift error is a realistic estimate of the actual redshift estimation error (see Fig.~\ref{fig:scalederrorhistogram}). While we observed redshift-dependent biases of up to $\Delta z=0.01$, \revision{the $z_{\mathrm{phot}} \pm \delta z_{\mathrm{phot}}$ confidence intervals provide a reasonably good approximation of the spectroscopic redshift (see Sec.~\ref{sec:accuracy})}. However, from $z \approx 0.6$, the coverage of our training set drops sharply, therefore so does the accuracy of our photometric redshifts.

In addition to the redshift error estimate, we provide further tools that allow users to select measurements of the desired accuracy. These include the photometric error class, the 3D redshift error map, and the bounding box volume of the nearest neighbours (see Sec.~\ref{sec:errordiscussion}, Sec.~\ref{sec:errormap} and Sec.~\ref{sec:practices}, respectively).

As opposed to purely empirical methods, our hybrid method fits a spectral template, which allows us to provide K-corrections and absolute magnitudes (see Sec.~\ref{sec:templatefit} and Sec.~\ref{sec:appendix1}).

In later releases, we intend to expand our training set as new data comes available, and also review our training set and methods with the intention of reducing biases and extending the useful coverage to higher redshifts. \revision{Having a less sparse sampling of high-redshift galaxies in photometric colour space would help reduce the pronounced negative bias in their redshift estimation.}


The photometric redshift database, corresponding documentation and tools are available online on the appropriate SDSS DR12 webpages.

\section*{Acknowledgments}

The realisation of this work was supported by the Hungarian OTKA NN grants 103244 and 114560.

\bibliographystyle{mn2e}

\begin{thebibliography}{66}
\expandafter\ifx\csname natexlab\endcsname\relax\def\natexlab#1{#1}\fi

\bibitem[{{Alam} {et~al}\mbox{.}(2015){Alam}, {Albareti}, {Allende Prieto},
  {Anders}, {Anderson}, {Anderton}, {Andrews}, {Armengaud}, {Aubourg},
  {Bailey}, \& et~al.}]{Alam2015}
{Alam} S. {et~al.}, 2015, \apjs, 219, 12

\bibitem[{{Arnouts} {et~al}\mbox{.}(2002){Arnouts}, {Moscardini}, {Vanzella},
  {Colombi}, {Cristiani}, {Fontana}, {Giallongo}, {Matarrese}, \&
  {Saracco}}]{Arnouts2002}
{Arnouts} S. {et~al.}, 2002, \mnras, 329, 355

\bibitem[{{Atek} {et~al}\mbox{.}(2011){Atek}, {Siana}, {Scarlata}, {Malkan},
  {McCarthy}, {Teplitz}, {Henry}, {Colbert}, {Bridge}, {Bunker}, {Dressler},
  {Fosbury}, {Hathi}, {Martin}, {Ross}, \& {Shim}}]{Atek2011}
{Atek} H. {et~al.}, 2011, \apj, 743, 121

\bibitem[{{Baldry} {et~al}\mbox{.}(2014){Baldry}, {Alpaslan}, {Bauer},
  {Bland-Hawthorn}, {Brough}, {Cluver}, {Croom}, {Davies}, {Driver},
  {Gunawardhana}, {Holwerda}, {Hopkins}, {Kelvin}, {Liske},
  {L{\'o}pez-S{\'a}nchez}, {Loveday}, {Norberg}, {Peacock}, {Robotham}, \&
  {Taylor}}]{Baldry2014}
{Baldry} I.~K. {et~al.}, 2014, \mnras, 441, 2440

\bibitem[{{Beck} {et~al}\mbox{.}(2016){Beck}, {Dobos}, {Yip}, {Szalay}, \&
  {Csabai}}]{Beck2016}
{Beck} R., {Dobos} L., {Yip} C.-W., {Szalay} A.~S., {Csabai} I., 2016, \mnras,
  457, 362

\bibitem[{{Ben{\'{\i}}tez}(2000)}]{Benitez2000}
{Ben{\'{\i}}tez} N., 2000, \apj, 536, 571

\bibitem[{{Bolzonella}, {Miralles} \& {Pell{\'o}}(2000){Bolzonella},
  {Miralles}, \& {Pell{\'o}}}]{Bolzonella2000}
{Bolzonella} M., {Miralles} J.-M., {Pell{\'o}} R., 2000, \aap, 363, 476

\bibitem[{{Brammer}, {van Dokkum} \& {Coppi}(2008){Brammer}, {van Dokkum}, \&
  {Coppi}}]{Brammer2008}
{Brammer} G.~B., {van Dokkum} P.~G., {Coppi} P., 2008, \apj, 686, 1503

\bibitem[{{Brescia} {et~al}\mbox{.}(2014){Brescia}, {Cavuoti}, {Longo}, \& {De
  Stefano}}]{Brescia2014}
{Brescia} M., {Cavuoti} S., {Longo} G., {De Stefano} V., 2014, \aap, 568, A126

\bibitem[{{Bruzual} \& {Charlot}(2003)}]{BruzualCharlot2003}
{Bruzual} G., {Charlot} S., 2003, \mnras, 344, 1000

\bibitem[{{Budav{\'a}ri}(2009)}]{Budavari2009}
{Budav{\'a}ri} T., 2009, \apj, 695, 747

\bibitem[{{Budav{\'a}ri} \& {Szalay}(2008)}]{Budavari2008}
{Budav{\'a}ri} T., {Szalay} A.~S., 2008, \apj, 679, 301

\bibitem[{{Budav{\'a}ri} {et~al}\mbox{.}(2000){Budav{\'a}ri}, {Szalay},
  {Connolly}, {Csabai}, \& {Dickinson}}]{Budavari2000}
{Budav{\'a}ri} T., {Szalay} A.~S., {Connolly} A.~J., {Csabai} I., {Dickinson}
  M., 2000, \aj, 120, 1588

\bibitem[{{Budav{\'a}ri} {et~al}\mbox{.}(2001){Budav{\'a}ri}, {Szalay},
  {Csabai}, {Connolly}, \& {Tsvetanov}}]{Budavari2001}
{Budav{\'a}ri} T., {Szalay} A.~S., {Csabai} I., {Connolly} A.~J., {Tsvetanov}
  Z., 2001, \aj, 121, 3266

\bibitem[{{Carliles} {et~al}\mbox{.}(2010){Carliles}, {Budav{\'a}ri}, {Heinis},
  {Priebe}, \& {Szalay}}]{Carliles2010}
{Carliles} S., {Budav{\'a}ri} T., {Heinis} S., {Priebe} C., {Szalay} A.~S.,
  2010, \apj, 712, 511

\bibitem[{{Coe} {et~al}\mbox{.}(2006){Coe}, {Ben{\'{\i}}tez}, {S{\'a}nchez},
  {Jee}, {Bouwens}, \& {Ford}}]{Coe2006}
{Coe} D., {Ben{\'{\i}}tez} N., {S{\'a}nchez} S.~F., {Jee} M., {Bouwens} R.,
  {Ford} H., 2006, \aj, 132, 926

\bibitem[{{Coil} {et~al}\mbox{.}(2011){Coil}, {Blanton}, {Burles}, {Cool},
  {Eisenstein}, {Moustakas}, {Wong}, {Zhu}, {Aird}, {Bernstein}, {Bolton}, \&
  {Hogg}}]{Coil2011}
{Coil} A.~L. {et~al.}, 2011, \apj, 741, 8

\bibitem[{{Coil} {et~al}\mbox{.}(2004){Coil}, {Newman}, {Kaiser}, {Davis},
  {Ma}, {Kocevski}, \& {Koo}}]{Coil2004}
{Coil} A.~L., {Newman} J.~A., {Kaiser} N., {Davis} M., {Ma} C.-P., {Kocevski}
  D.~D., {Koo} D.~C., 2004, \apj, 617, 765

\bibitem[{{Colless} {et~al}\mbox{.}(2001){Colless}, {Dalton}, {Maddox},
  {Sutherland}, {Norberg}, {Cole}, {Bland-Hawthorn}, {Bridges}, {Cannon},
  {Collins}, {Couch}, {Cross}, {Deeley}, {De Propris}, {Driver}, {Efstathiou},
  {Ellis}, {Frenk}, {Glazebrook}, {Jackson}, {Lahav}, {Lewis}, {Lumsden},
  {Madgwick}, {Peacock}, {Peterson}, {Price}, {Seaborne}, \&
  {Taylor}}]{Colless2001}
{Colless} M. {et~al.}, 2001, \mnras, 328, 1039

\bibitem[{{Colless} {et~al}\mbox{.}(2003){Colless}, {Peterson}, {Jackson},
  {Peacock}, {Cole}, {Norberg}, {Baldry}, {Baugh}, {Bland-Hawthorn}, {Bridges},
  {Cannon}, {Collins}, {Couch}, {Cross}, {Dalton}, {De Propris}, {Driver},
  {Efstathiou}, {Ellis}, {Frenk}, {Glazebrook}, {Lahav}, {Lewis}, {Lumsden},
  {Maddox}, {Madgwick}, {Sutherland}, \& {Taylor}}]{Colless2003}
{Colless} M. {et~al.}, 2003, ArXiv Astrophysics e-prints

\bibitem[{{Collister} {et~al}\mbox{.}(2007){Collister}, {Lahav}, {Blake},
  {Cannon}, {Croom}, {Drinkwater}, {Edge}, {Eisenstein}, {Loveday}, {Nichol},
  {Pimbblet}, {de Propris}, {Roseboom}, {Ross}, {Schneider}, {Shanks}, \&
  {Wake}}]{Collister2007}
{Collister} A. {et~al.}, 2007, \mnras, 375, 68

\bibitem[{{Cool} {et~al}\mbox{.}(2013){Cool}, {Moustakas}, {Blanton}, {Burles},
  {Coil}, {Eisenstein}, {Wong}, {Zhu}, {Aird}, {Bernstein}, {Bolton}, {Hogg},
  \& {Mendez}}]{Cool2013}
{Cool} R.~J. {et~al.}, 2013, \apj, 767, 118

\bibitem[{{Csabai} {et~al}\mbox{.}(2003){Csabai}, {Budav{\'a}ri}, {Connolly},
  {Szalay}, {Gy{\H o}ry}, {Ben{\'{\i}}tez}, {Annis}, {Brinkmann}, {Eisenstein},
  {Fukugita}, {Gunn}, {Kent}, {Lupton}, {Nichol}, \& {Stoughton}}]{Csabai2003}
{Csabai} I. {et~al.}, 2003, \aj, 125, 580

\bibitem[{{Csabai} {et~al}\mbox{.}(2000){Csabai}, {Connolly}, {Szalay}, \&
  {Budav{\'a}ri}}]{Csabai2000}
{Csabai} I., {Connolly} A.~J., {Szalay} A.~S., {Budav{\'a}ri} T., 2000, \aj,
  119, 69

\bibitem[{{Csabai} {et~al}\mbox{.}(2007){Csabai}, {Dobos}, {Trencs{\'e}ni},
  {Herczegh}, {J{\'o}zsa}, {Purger}, {Budav{\'a}ri}, \& {Szalay}}]{Csabai2007}
{Csabai} I., {Dobos} L., {Trencs{\'e}ni} M., {Herczegh} G., {J{\'o}zsa} P.,
  {Purger} N., {Budav{\'a}ri} T., {Szalay} A.~S., 2007, Astronomische
  Nachrichten, 328, 852

\bibitem[{{Davis} {et~al}\mbox{.}(2003){Davis}, {Faber}, {Newman}, {Phillips},
  {Ellis}, {Steidel}, {Conselice}, {Coil}, {Finkbeiner}, {Koo}, {Guhathakurta},
  {Weiner}, {Schiavon}, {Willmer}, {Kaiser}, {Luppino}, {Wirth}, {Connolly},
  {Eisenhardt}, {Cooper}, \& {Gerke}}]{Davis2003}
{Davis} M. {et~al.}, 2003, in Society of Photo-Optical Instrumentation
  Engineers (SPIE) Conference Series, Vol. 4834, Discoveries and Research
  Prospects from 6- to 10-Meter-Class Telescopes II, {Guhathakurta} P., ed.,
  pp. 161--172

\bibitem[{{Dawson} {et~al}\mbox{.}(2013){Dawson}, {Schlegel}, {Ahn},
  {Anderson}, {Aubourg}, {Bailey}, {Barkhouser}, {Bautista}, {Beifiori},
  {Berlind}, {Bhardwaj}, {Bizyaev}, {Blake}, {Blanton}, {Blomqvist}, {Bolton},
  {Borde}, {Bovy}, {Brandt}, {Brewington}, {Brinkmann}, {Brown}, {Brownstein},
  {Bundy}, {Busca}, {Carithers}, {Carnero}, {Carr}, {Chen}, {Comparat},
  {Connolly}, {Cope}, {Croft}, {Cuesta}, {da Costa}, {Davenport}, {Delubac},
  {de Putter}, {Dhital}, {Ealet}, {Ebelke}, {Eisenstein}, {Escoffier}, {Fan},
  {Filiz Ak}, {Finley}, {Font-Ribera}, {G{\'e}nova-Santos}, {Gunn}, {Guo},
  {Haggard}, {Hall}, {Hamilton}, {Harris}, {Harris}, {Ho}, {Hogg}, {Holder},
  {Honscheid}, {Huehnerhoff}, {Jordan}, {Jordan}, {Kauffmann}, {Kazin},
  {Kirkby}, {Klaene}, {Kneib}, {Le Goff}, {Lee}, {Long}, {Loomis}, {Lundgren},
  {Lupton}, {Maia}, {Makler}, {Malanushenko}, {Malanushenko}, {Mandelbaum},
  {Manera}, {Maraston}, {Margala}, {Masters}, {McBride}, {McDonald}, {McGreer},
  {McMahon}, {Mena}, {Miralda-Escud{\'e}}, {Montero-Dorta}, {Montesano},
  {Muna}, {Myers}, {Naugle}, {Nichol}, {Noterdaeme}, {Nuza}, {Olmstead},
  {Oravetz}, {Oravetz}, {Owen}, {Padmanabhan}, {Palanque-Delabrouille}, {Pan},
  {Parejko}, {P{\^a}ris}, {Percival}, {P{\'e}rez-Fournon},
  {P{\'e}rez-R{\`a}fols}, {Petitjean}, {Pfaffenberger}, {Pforr}, {Pieri},
  {Prada}, {Price-Whelan}, {Raddick}, {Rebolo}, {Rich}, {Richards}, {Rockosi},
  {Roe}, {Ross}, {Ross}, {Rossi}, {Rubi{\~n}o-Martin}, {Samushia},
  {S{\'a}nchez}, {Sayres}, {Schmidt}, {Schneider}, {Sc{\'o}ccola}, {Seo},
  {Shelden}, {Sheldon}, {Shen}, {Shu}, {Slosar}, {Smee}, {Snedden}, {Stauffer},
  {Steele}, {Strauss}, {Streblyanska}, {Suzuki}, {Swanson}, {Tal}, {Tanaka},
  {Thomas}, {Tinker}, {Tojeiro}, {Tremonti}, {Vargas Maga{\~n}a}, {Verde},
  {Viel}, {Wake}, {Watson}, {Weaver}, {Weinberg}, {Weiner}, {West}, {White},
  {Wood-Vasey}, {Yeche}, {Zehavi}, {Zhao}, \& {Zheng}}]{Dawson2013}
{Dawson} K.~S. {et~al.}, 2013, \aj, 145, 10

\bibitem[{{Dobos} {et~al}\mbox{.}(2012){Dobos}, {Csabai}, {Yip},
  {Budav{\'a}ri}, {Wild}, \& {Szalay}}]{Dobos2012}
{Dobos} L., {Csabai} I., {Yip} C.-W., {Budav{\'a}ri} T., {Wild} V., {Szalay}
  A.~S., 2012, \mnras, 420, 1217

\bibitem[{{Doi} {et~al}\mbox{.}(2010){Doi}, {Tanaka}, {Fukugita}, {Gunn},
  {Yasuda}, {Ivezi{\'c}}, {Brinkmann}, {de Haars}, {Kleinman}, {Krzesinski}, \&
  {French Leger}}]{Doi2010}
{Doi} M. {et~al.}, 2010, \aj, 139, 1628

\bibitem[{{Drinkwater} {et~al}\mbox{.}(2010){Drinkwater}, {Jurek}, {Blake},
  {Woods}, {Pimbblet}, {Glazebrook}, {Sharp}, {Pracy}, {Brough}, {Colless},
  {Couch}, {Croom}, {Davis}, {Forbes}, {Forster}, {Gilbank}, {Gladders},
  {Jelliffe}, {Jones}, {Li}, {Madore}, {Martin}, {Poole}, {Small}, {Wisnioski},
  {Wyder}, \& {Yee}}]{Drinkwater2010}
{Drinkwater} M.~J. {et~al.}, 2010, \mnras, 401, 1429

\bibitem[{{Driver} {et~al}\mbox{.}(2011){Driver}, {Hill}, {Kelvin}, {Robotham},
  {Liske}, {Norberg}, {Baldry}, {Bamford}, {Hopkins}, {Loveday}, {Peacock},
  {Andrae}, {Bland-Hawthorn}, {Brough}, {Brown}, {Cameron}, {Ching}, {Colless},
  {Conselice}, {Croom}, {Cross}, {de Propris}, {Dye}, {Drinkwater}, {Ellis},
  {Graham}, {Grootes}, {Gunawardhana}, {Jones}, {van Kampen}, {Maraston},
  {Nichol}, {Parkinson}, {Phillipps}, {Pimbblet}, {Popescu}, {Prescott},
  {Roseboom}, {Sadler}, {Sansom}, {Sharp}, {Smith}, {Taylor}, {Thomas},
  {Tuffs}, {Wijesinghe}, {Dunne}, {Frenk}, {Jarvis}, {Madore}, {Meyer},
  {Seibert}, {Staveley-Smith}, {Sutherland}, \& {Warren}}]{Driver2011}
{Driver} S.~P. {et~al.}, 2011, \mnras, 413, 971

\bibitem[{{Eisenstein} {et~al}\mbox{.}(2001){Eisenstein}, {Annis}, {Gunn},
  {Szalay}, {Connolly}, {Nichol}, {Bahcall}, {Bernardi}, {Burles}, {Castander},
  {Fukugita}, {Hogg}, {Ivezi{\'c}}, {Knapp}, {Lupton}, {Narayanan}, {Postman},
  {Reichart}, {Richmond}, {Schneider}, {Schlegel}, {Strauss}, {SubbaRao},
  {Tucker}, {Vanden Berk}, {Vogeley}, {Weinberg}, \& {Yanny}}]{Eisenstein2001}
{Eisenstein} D.~J. {et~al.}, 2001, \aj, 122, 2267

\bibitem[{{Eisenstein} {et~al}\mbox{.}(2011){Eisenstein}, {Weinberg}, {Agol},
  {Aihara}, {Allende Prieto}, {Anderson}, {Arns}, {Aubourg}, {Bailey},
  {Balbinot}, \& et~al.}]{Eisenstein2011}
{Eisenstein} D.~J. {et~al.}, 2011, \aj, 142, 72

\bibitem[{{Feldmann} {et~al}\mbox{.}(2006){Feldmann}, {Carollo}, {Porciani},
  {Lilly}, {Capak}, {Taniguchi}, {Le F{\`e}vre}, {Renzini}, {Scoville},
  {Ajiki}, {Aussel}, {Contini}, {McCracken}, {Mobasher}, {Murayama}, {Sanders},
  {Sasaki}, {Scarlata}, {Scodeggio}, {Shioya}, {Silverman}, {Takahashi},
  {Thompson}, \& {Zamorani}}]{Feldmann2006}
{Feldmann} R. {et~al.}, 2006, \mnras, 372, 565

\bibitem[{{Ferland} {et~al}\mbox{.}(2013){Ferland}, {Porter}, {van Hoof},
  {Williams}, {Abel}, {Lykins}, {Shaw}, {Henney}, \& {Stancil}}]{cloudy}
{Ferland} G.~J. {et~al.}, 2013, Revista Mexicana de Astronom\'{i}a y
  Astrof\'{i}sica, 49, 137

\bibitem[{{Fioc} \& {Rocca-Volmerange}(1997)}]{Pegase1}
{Fioc} M., {Rocca-Volmerange} B., 1997, \aap, 326, 950

\bibitem[{{Garilli} {et~al}\mbox{.}(2014){Garilli}, {Guzzo}, {Scodeggio},
  {Bolzonella}, {Abbas}, {Adami}, {Arnouts}, {Bel}, {Bottini}, {Branchini},
  {Cappi}, {Coupon}, {Cucciati}, {Davidzon}, {De Lucia}, {de la Torre},
  {Franzetti}, {Fritz}, {Fumana}, {Granett}, {Ilbert}, {Iovino}, {Krywult}, {Le
  Brun}, {Le F{\`e}vre}, {Maccagni}, {Ma{\l}ek}, {Marulli}, {McCracken},
  {Paioro}, {Polletta}, {Pollo}, {Schlagenhaufer}, {Tasca}, {Tojeiro},
  {Vergani}, {Zamorani}, {Zanichelli}, {Burden}, {Di Porto}, {Marchetti},
  {Marinoni}, {Mellier}, {Moscardini}, {Nichol}, {Peacock}, {Percival},
  {Phleps}, \& {Wolk}}]{Garilli2014}
{Garilli} B. {et~al.}, 2014, \aap, 562, A23

\bibitem[{{Garilli} {et~al}\mbox{.}(2008){Garilli}, {Le F{\`e}vre}, {Guzzo},
  {Maccagni}, {Le Brun}, {de la Torre}, {Meneux}, {Tresse}, {Franzetti},
  {Zamorani}, {Zanichelli}, {Gregorini}, {Vergani}, {Bottini}, {Scaramella},
  {Scodeggio}, {Vettolani}, {Adami}, {Arnouts}, {Bardelli}, {Bolzonella},
  {Cappi}, {Charlot}, {Ciliegi}, {Contini}, {Foucaud}, {Gavignaud}, {Ilbert},
  {Iovino}, {Lamareille}, {McCracken}, {Marano}, {Marinoni}, {Mazure},
  {Merighi}, {Paltani}, {Pell{\`o}}, {Pollo}, {Pozzetti}, {Radovich}, {Zucca},
  {Blaizot}, {Bongiorno}, {Cucciati}, {Mellier}, {Moreau}, \&
  {Paioro}}]{Garilli2008}
{Garilli} B. {et~al.}, 2008, \aap, 486, 683

\bibitem[{{Gerdes} {et~al}\mbox{.}(2010){Gerdes}, {Sypniewski}, {McKay}, {Hao},
  {Weis}, {Wechsler}, \& {Busha}}]{Gerdes2010}
{Gerdes} D.~W., {Sypniewski} A.~J., {McKay} T.~A., {Hao} J., {Weis} M.~R.,
  {Wechsler} R.~H., {Busha} M.~T., 2010, \apj, 715, 823

\bibitem[{{Gunn} {et~al}\mbox{.}(1998){Gunn}, {Carr}, {Rockosi}, {Sekiguchi},
  {Berry}, {Elms}, {de Haas}, {Ivezi{\'c}}, {Knapp}, {Lupton}, {Pauls},
  {Simcoe}, {Hirsch}, {Sanford}, {Wang}, {York}, {Harris}, {Annis}, {Bartozek},
  {Boroski}, {Bakken}, {Haldeman}, {Kent}, {Holm}, {Holmgren}, {Petravick},
  {Prosapio}, {Rechenmacher}, {Doi}, {Fukugita}, {Shimasaku}, {Okada}, {Hull},
  {Siegmund}, {Mannery}, {Blouke}, {Heidtman}, {Schneider}, {Lucinio}, \&
  {Brinkman}}]{Gunn1998}
{Gunn} J.~E. {et~al.}, 1998, \aj, 116, 3040

\bibitem[{{Guzzo} {et~al}\mbox{.}(2014){Guzzo}, {Scodeggio}, {Garilli},
  {Granett}, {Fritz}, {Abbas}, {Adami}, {Arnouts}, {Bel}, {Bolzonella},
  {Bottini}, {Branchini}, {Cappi}, {Coupon}, {Cucciati}, {Davidzon}, {De
  Lucia}, {de la Torre}, {Franzetti}, {Fumana}, {Hudelot}, {Ilbert}, {Iovino},
  {Krywult}, {Le Brun}, {Le F{\`e}vre}, {Maccagni}, {Ma{\l}ek}, {Marulli},
  {McCracken}, {Paioro}, {Peacock}, {Polletta}, {Pollo}, {Schlagenhaufer},
  {Tasca}, {Tojeiro}, {Vergani}, {Zamorani}, {Zanichelli}, {Burden}, {Di
  Porto}, {Marchetti}, {Marinoni}, {Mellier}, {Moscardini}, {Nichol},
  {Percival}, {Phleps}, \& {Wolk}}]{Guzzo2014}
{Guzzo} L. {et~al.}, 2014, \aap, 566, A108

\bibitem[{{Gy{\H o}ry} {et~al}\mbox{.}(2011){Gy{\H o}ry}, {Szalay},
  {Budav{\'a}ri}, {Csabai}, \& {Charlot}}]{Gyory2011}
{Gy{\H o}ry} Z., {Szalay} A.~S., {Budav{\'a}ri} T., {Csabai} I., {Charlot} S.,
  2011, \aj, 141, 133

\bibitem[{{Hinshaw} {et~al}\mbox{.}(2009){Hinshaw}, {Weiland}, {Hill},
  {Odegard}, {Larson}, {Bennett}, {Dunkley}, {Gold}, {Greason}, {Jarosik},
  {Komatsu}, {Nolta}, {Page}, {Spergel}, {Wollack}, {Halpern}, {Kogut},
  {Limon}, {Meyer}, {Tucker}, \& {Wright}}]{Hinshaw2009}
{Hinshaw} G. {et~al.}, 2009, \apjs, 180, 225

\bibitem[{{Ilbert} {et~al}\mbox{.}(2006){Ilbert}, {Arnouts}, {McCracken},
  {Bolzonella}, {Bertin}, {Le F{\`e}vre}, {Mellier}, {Zamorani}, {Pell{\`o}},
  {Iovino}, {Tresse}, {Le Brun}, {Bottini}, {Garilli}, {Maccagni}, {Picat},
  {Scaramella}, {Scodeggio}, {Vettolani}, {Zanichelli}, {Adami}, {Bardelli},
  {Cappi}, {Charlot}, {Ciliegi}, {Contini}, {Cucciati}, {Foucaud}, {Franzetti},
  {Gavignaud}, {Guzzo}, {Marano}, {Marinoni}, {Mazure}, {Meneux}, {Merighi},
  {Paltani}, {Pollo}, {Pozzetti}, {Radovich}, {Zucca}, {Bondi}, {Bongiorno},
  {Busarello}, {de La Torre}, {Gregorini}, {Lamareille}, {Mathez}, {Merluzzi},
  {Ripepi}, {Rizzo}, \& {Vergani}}]{Ilbert2006}
{Ilbert} O. {et~al.}, 2006, \aap, 457, 841

\bibitem[{{Ivezic} {et~al}\mbox{.}(2008){Ivezic}, {Tyson}, {Abel}, {Acosta},
  {Allsman}, {AlSayyad}, {Anderson}, {Andrew}, {Angel}, {Angeli}, {Ansari},
  {Antilogus}, {Arndt}, {Astier}, {Aubourg}, {Axelrod}, {Bard}, {Barr},
  {Barrau}, {Bartlett}, {Bauman}, {Beaumont}, {Becker}, {Becla}, {Beldica},
  {Bellavia}, {Blanc}, {Blandford}, {Bloom}, {Bogart}, {Borne}, {Bosch},
  {Boutigny}, {Brandt}, {Brown}, {Bullock}, {Burchat}, {Burke}, {Cagnoli},
  {Calabrese}, {Chandrasekharan}, {Chesley}, {Cheu}, {Chiang}, {Claver},
  {Connolly}, {Cook}, {Cooray}, {Covey}, {Cribbs}, {Cui}, {Cutri}, {Daubard},
  {Daues}, {Delgado}, {Digel}, {Doherty}, {Dubois}, {Dubois-Felsmann},
  {Durech}, {Eracleous}, {Ferguson}, {Frank}, {Freemon}, {Gangler}, {Gawiser},
  {Geary}, {Gee}, {Geha}, {Gibson}, {Gilmore}, {Glanzman}, {Goodenow},
  {Gressler}, {Gris}, {Guyonnet}, {Hascall}, {Haupt}, {Hernandez}, {Hogan},
  {Huang}, {Huffer}, {Innes}, {Jacoby}, {Jain}, {Jee}, {Jernigan},
  {Jevremovic}, {Johns}, {Jones}, {Juramy-Gilles}, {Juric}, {Kahn}, {Kalirai},
  {Kallivayalil}, {Kalmbach}, {Kantor}, {Kasliwal}, {Kessler}, {Kirkby},
  {Knox}, {Kotov}, {Krabbendam}, {Krughoff}, {Kubanek}, {Kuczewski},
  {Kulkarni}, {Lambert}, {Le Guillou}, {Levine}, {Liang}, {Lim}, {Lintott},
  {Lupton}, {Mahabal}, {Marshall}, {Marshall}, {May}, {McKercher}, {Migliore},
  {Miller}, {Mills}, {Monet}, {Moniez}, {Neill}, {Nief}, {Nomerotski},
  {Nordby}, {O'Connor}, {Oliver}, {Olivier}, {Olsen}, {Ortiz}, {Owen}, {Pain},
  {Peterson}, {Petry}, {Pierfederici}, {Pietrowicz}, {Pike}, {Pinto}, {Plante},
  {Plate}, {Price}, {Prouza}, {Radeka}, {Rajagopal}, {Rasmussen}, {Regnault},
  {Ridgway}, {Ritz}, {Rosing}, {Roucelle}, {Rumore}, {Russo}, {Saha},
  {Sassolas}, {Schalk}, {Schindler}, {Schneider}, {Schumacher}, {Sebag},
  {Sembroski}, {Seppala}, {Shipsey}, {Silvestri}, {Smith}, {Smith}, {Strauss},
  {Stubbs}, {Sweeney}, {Szalay}, {Takacs}, {Thaler}, {Van Berg}, {Vanden Berk},
  {Vetter}, {Virieux}, {Xin}, {Walkowicz}, {Walter}, {Wang}, {Warner},
  {Willman}, {Wittman}, {Wolff}, {Wood-Vasey}, {Yoachim}, {Zhan}, \& {for the
  LSST Collaboration}}]{Ivezic2008}
{Ivezic} Z. {et~al.}, 2008, ArXiv e-prints

\bibitem[{{Jones} {et~al}\mbox{.}(2009){Jones}, {Read}, {Saunders}, {Colless},
  {Jarrett}, {Parker}, {Fairall}, {Mauch}, {Sadler}, {Watson}, {Burton},
  {Campbell}, {Cass}, {Croom}, {Dawe}, {Fiegert}, {Frankcombe}, {Hartley},
  {Huchra}, {James}, {Kirby}, {Lahav}, {Lucey}, {Mamon}, {Moore}, {Peterson},
  {Prior}, {Proust}, {Russell}, {Safouris}, {Wakamatsu}, {Westra}, \&
  {Williams}}]{Jones2009}
{Jones} D.~H. {et~al.}, 2009, \mnras, 399, 683

\bibitem[{{Jones} {et~al}\mbox{.}(2004){Jones}, {Saunders}, {Colless}, {Read},
  {Parker}, {Watson}, {Campbell}, {Burkey}, {Mauch}, {Moore}, {Hartley},
  {Cass}, {James}, {Russell}, {Fiegert}, {Dawe}, {Huchra}, {Jarrett}, {Lahav},
  {Lucey}, {Mamon}, {Proust}, {Sadler}, \& {Wakamatsu}}]{Jones2004}
{Jones} D.~H. {et~al.}, 2004, \mnras, 355, 747

\bibitem[{{Le F{\`e}vre} {et~al}\mbox{.}(2004){Le F{\`e}vre}, {Mellier},
  {McCracken}, {Foucaud}, {Gwyn}, {Radovich}, {Dantel-Fort}, {Bertin},
  {Moreau}, {Cuillandre}, {Pierre}, {Le Brun}, {Mazure}, \&
  {Tresse}}]{LeFevre2004}
{Le F{\`e}vre} O. {et~al.}, 2004, \aap, 417, 839

\bibitem[{{Lilly} {et~al}\mbox{.}(2009){Lilly}, {Le Brun}, {Maier}, {Mainieri},
  {Mignoli}, {Scodeggio}, {Zamorani}, {Carollo}, {Contini}, {Kneib}, {Le
  F{\`e}vre}, {Renzini}, {Bardelli}, {Bolzonella}, {Bongiorno}, {Caputi},
  {Coppa}, {Cucciati}, {de la Torre}, {de Ravel}, {Franzetti}, {Garilli},
  {Iovino}, {Kampczyk}, {Kovac}, {Knobel}, {Lamareille}, {Le Borgne}, {Pello},
  {Peng}, {P{\'e}rez-Montero}, {Ricciardelli}, {Silverman}, {Tanaka}, {Tasca},
  {Tresse}, {Vergani}, {Zucca}, {Ilbert}, {Salvato}, {Oesch}, {Abbas},
  {Bottini}, {Capak}, {Cappi}, {Cassata}, {Cimatti}, {Elvis}, {Fumana},
  {Guzzo}, {Hasinger}, {Koekemoer}, {Leauthaud}, {Maccagni}, {Marinoni},
  {McCracken}, {Memeo}, {Meneux}, {Porciani}, {Pozzetti}, {Sanders},
  {Scaramella}, {Scarlata}, {Scoville}, {Shopbell}, \& {Taniguchi}}]{Lilly2009}
{Lilly} S.~J. {et~al.}, 2009, \apjs, 184, 218

\bibitem[{{Lilly} {et~al}\mbox{.}(2007){Lilly}, {Le F{\`e}vre}, {Renzini},
  {Zamorani}, {Scodeggio}, {Contini}, {Carollo}, {Hasinger}, {Kneib}, {Iovino},
  {Le Brun}, {Maier}, {Mainieri}, {Mignoli}, {Silverman}, {Tasca},
  {Bolzonella}, {Bongiorno}, {Bottini}, {Capak}, {Caputi}, {Cimatti},
  {Cucciati}, {Daddi}, {Feldmann}, {Franzetti}, {Garilli}, {Guzzo}, {Ilbert},
  {Kampczyk}, {Kovac}, {Lamareille}, {Leauthaud}, {Borgne}, {McCracken},
  {Marinoni}, {Pello}, {Ricciardelli}, {Scarlata}, {Vergani}, {Sanders},
  {Schinnerer}, {Scoville}, {Taniguchi}, {Arnouts}, {Aussel}, {Bardelli},
  {Brusa}, {Cappi}, {Ciliegi}, {Finoguenov}, {Foucaud}, {Franceschini},
  {Halliday}, {Impey}, {Knobel}, {Koekemoer}, {Kurk}, {Maccagni}, {Maddox},
  {Marano}, {Marconi}, {Meneux}, {Mobasher}, {Moreau}, {Peacock}, {Porciani},
  {Pozzetti}, {Scaramella}, {Schiminovich}, {Shopbell}, {Smail}, {Thompson},
  {Tresse}, {Vettolani}, {Zanichelli}, \& {Zucca}}]{Lilly2007}
{Lilly} S.~J. {et~al.}, 2007, \apjs, 172, 70

\bibitem[{{Lupton}, {Gunn} \& {Szalay}(1999){Lupton}, {Gunn}, \&
  {Szalay}}]{Lupton1999}
{Lupton} R.~H., {Gunn} J.~E., {Szalay} A.~S., 1999, \aj, 118, 1406

\bibitem[{{Maraston} \& {Str{\"o}mb{\"a}ck}(2011)}]{Maraston2011}
{Maraston} C., {Str{\"o}mb{\"a}ck} G., 2011, \mnras, 418, 2785

\bibitem[{{Marchetti} {et~al}\mbox{.}(2013){Marchetti}, {Granett}, {Guzzo},
  {Fritz}, {Garilli}, {Scodeggio}, {Abbas}, {Adami}, {Arnouts}, {Bolzonella},
  {Bottini}, {Cappi}, {Coupon}, {Cucciati}, {De Lucia}, {de la Torre},
  {Franzetti}, {Fumana}, {Ilbert}, {Iovino}, {Krywult}, {Le Brun}, {Le Fevre},
  {Maccagni}, {Malek}, {Marulli}, {McCracken}, {Meneux}, {Paioro}, {Polletta},
  {Pollo}, {Schlagenhaufer}, {Tasca}, {Tojeiro}, {Vergani}, {Zanichelli},
  {Bel}, {Bersanelli}, {Blaizot}, {Branchini}, {Burden}, {Davidzon}, {Di
  Porto}, {Guennou}, {Marinoni}, {Mellier}, {Moscardini}, {Nichol}, {Peacock},
  {Percival}, {Phleps}, {Schimd}, {Wolk}, \& {Zamorani}}]{Marchetti2013}
{Marchetti} A. {et~al.}, 2013, \mnras, 428, 1424

\bibitem[{{Newman} {et~al}\mbox{.}(2013){Newman}, {Cooper}, {Davis}, {Faber},
  {Coil}, {Guhathakurta}, {Koo}, {Phillips}, {Conroy}, {Dutton}, {Finkbeiner},
  {Gerke}, {Rosario}, {Weiner}, {Willmer}, {Yan}, {Harker}, {Kassin},
  {Konidaris}, {Lai}, {Madgwick}, {Noeske}, {Wirth}, {Connolly}, {Kaiser},
  {Kirby}, {Lemaux}, {Lin}, {Lotz}, {Luppino}, {Marinoni}, {Matthews},
  {Metevier}, \& {Schiavon}}]{Newman2013}
{Newman} J.~A. {et~al.}, 2013, \apjs, 208, 5

\bibitem[{{Parkinson} {et~al}\mbox{.}(2012){Parkinson}, {Riemer-S{\o}rensen},
  {Blake}, {Poole}, {Davis}, {Brough}, {Colless}, {Contreras}, {Couch},
  {Croom}, {Croton}, {Drinkwater}, {Forster}, {Gilbank}, {Gladders},
  {Glazebrook}, {Jelliffe}, {Jurek}, {Li}, {Madore}, {Martin}, {Pimbblet},
  {Pracy}, {Sharp}, {Wisnioski}, {Woods}, {Wyder}, \& {Yee}}]{Parkinson2012}
{Parkinson} D. {et~al.}, 2012, \prd, 86, 103518

\bibitem[{{Reis} {et~al}\mbox{.}(2012){Reis}, {Soares-Santos}, {Annis},
  {Dodelson}, {Hao}, {Johnston}, {Kubo}, {Lin}, {Seo}, \& {Simet}}]{Reis2012}
{Reis} R.~R.~R. {et~al.}, 2012, \apj, 747, 59

\bibitem[{{Schlegel}, {Finkbeiner} \& {Davis}(1998){Schlegel}, {Finkbeiner}, \&
  {Davis}}]{Schlegel1998}
{Schlegel} D.~J., {Finkbeiner} D.~P., {Davis} M., 1998, \apj, 500, 525

\bibitem[{{Scranton} {et~al}\mbox{.}(2005){Scranton}, {Connolly}, {Szalay},
  {Lupton}, {Johnston}, {Budavari}, {Brinkman}, \& {Fukugita}}]{Scranton2005}
{Scranton} R., {Connolly} A.~J., {Szalay} A.~S., {Lupton} R.~H., {Johnston} D.,
  {Budavari} T., {Brinkman} J., {Fukugita} M., 2005, ArXiv Astrophysics
  e-prints

\bibitem[{{Smee} {et~al}\mbox{.}(2013){Smee}, {Gunn}, {Uomoto}, {Roe},
  {Schlegel}, {Rockosi}, {Carr}, {Leger}, {Dawson}, {Olmstead}, {Brinkmann},
  {Owen}, {Barkhouser}, {Honscheid}, {Harding}, {Long}, {Lupton}, {Loomis},
  {Anderson}, {Annis}, {Bernardi}, {Bhardwaj}, {Bizyaev}, {Bolton},
  {Brewington}, {Briggs}, {Burles}, {Burns}, {Castander}, {Connolly},
  {Davenport}, {Ebelke}, {Epps}, {Feldman}, {Friedman}, {Frieman}, {Heckman},
  {Hull}, {Knapp}, {Lawrence}, {Loveday}, {Mannery}, {Malanushenko},
  {Malanushenko}, {Merrelli}, {Muna}, {Newman}, {Nichol}, {Oravetz}, {Pan},
  {Pope}, {Ricketts}, {Shelden}, {Sandford}, {Siegmund}, {Simmons}, {Smith},
  {Snedden}, {Schneider}, {SubbaRao}, {Tremonti}, {Waddell}, \&
  {York}}]{Smee2013}
{Smee} S.~A. {et~al.}, 2013, \aj, 146, 32

\bibitem[{{Stasi{\'n}ska}(1984)}]{Stasinska1984}
{Stasi{\'n}ska} G., 1984, \aaps, 55, 15

\bibitem[{{Strauss} {et~al}\mbox{.}(2002){Strauss}, {Weinberg}, {Lupton},
  {Narayanan}, {Annis}, {Bernardi}, {Blanton}, {Burles}, {Connolly},
  {Dalcanton}, {Doi}, {Eisenstein}, {Frieman}, {Fukugita}, {Gunn},
  {Ivezi{\'c}}, {Kent}, {Kim}, {Knapp}, {Kron}, {Munn}, {Newberg}, {Nichol},
  {Okamura}, {Quinn}, {Richmond}, {Schlegel}, {Shimasaku}, {SubbaRao},
  {Szalay}, {Vanden Berk}, {Vogeley}, {Yanny}, {Yasuda}, {York}, \&
  {Zehavi}}]{Strauss2002}
{Strauss} M.~A. {et~al.}, 2002, \aj, 124, 1810

\bibitem[{{Tonry} {et~al}\mbox{.}(2012){Tonry}, {Stubbs}, {Lykke}, {Doherty},
  {Shivvers}, {Burgett}, {Chambers}, {Hodapp}, {Kaiser}, {Kudritzki},
  {Magnier}, {Morgan}, {Price}, \& {Wainscoat}}]{Tonry2012}
{Tonry} J.~L. {et~al.}, 2012, \apj, 750, 99

\bibitem[{{Vazdekis} {et~al}\mbox{.}(2012){Vazdekis}, {Ricciardelli},
  {Cenarro}, {Rivero-Gonz{\'a}lez}, {D{\'{\i}}az-Garc{\'{\i}}a}, \&
  {Falc{\'o}n-Barroso}}]{Vazdekis2012}
{Vazdekis} A., {Ricciardelli} E., {Cenarro} A.~J., {Rivero-Gonz{\'a}lez} J.~G.,
  {D{\'{\i}}az-Garc{\'{\i}}a} L.~A., {Falc{\'o}n-Barroso} J., 2012, \mnras,
  424, 157

\bibitem[{{Wolf} {et~al}\mbox{.}(2003){Wolf}, {Meisenheimer}, {Rix}, {Borch},
  {Dye}, \& {Kleinheinrich}}]{Wolf2003}
{Wolf} C., {Meisenheimer} K., {Rix} H.-W., {Borch} A., {Dye} S.,
  {Kleinheinrich} M., 2003, \aap, 401, 73

\bibitem[{{Yip} {et~al}\mbox{.}(2004){Yip}, {Connolly}, {Szalay},
  {Budav{\'a}ri}, {SubbaRao}, {Frieman}, {Nichol}, {Hopkins}, {York},
  {Okamura}, {Brinkmann}, {Csabai}, {Thakar}, {Fukugita}, \&
  {Ivezi{\'c}}}]{Yip2004}
{Yip} C.~W. {et~al.}, 2004, \aj, 128, 585

\bibitem[{{York} {et~al}\mbox{.}(2000){York}, {Adelman}, {Anderson},
  {Anderson}, {Annis}, {Bahcall}, {Bakken}, {Barkhouser}, {Bastian}, {Berman},
  {Boroski}, {Bracker}, {Briegel}, {Briggs}, {Brinkmann}, {Brunner}, {Burles},
  {Carey}, {Carr}, {Castander}, {Chen}, {Colestock}, {Connolly}, {Crocker},
  {Csabai}, {Czarapata}, {Davis}, {Doi}, {Dombeck}, {Eisenstein}, {Ellman},
  {Elms}, {Evans}, {Fan}, {Federwitz}, {Fiscelli}, {Friedman}, {Frieman},
  {Fukugita}, {Gillespie}, {Gunn}, {Gurbani}, {de Haas}, {Haldeman}, {Harris},
  {Hayes}, {Heckman}, {Hennessy}, {Hindsley}, {Holm}, {Holmgren}, {Huang},
  {Hull}, {Husby}, {Ichikawa}, {Ichikawa}, {Ivezi{\'c}}, {Kent}, {Kim},
  {Kinney}, {Klaene}, {Kleinman}, {Kleinman}, {Knapp}, {Korienek}, {Kron},
  {Kunszt}, {Lamb}, {Lee}, {Leger}, {Limmongkol}, {Lindenmeyer}, {Long},
  {Loomis}, {Loveday}, {Lucinio}, {Lupton}, {MacKinnon}, {Mannery}, {Mantsch},
  {Margon}, {McGehee}, {McKay}, {Meiksin}, {Merelli}, {Monet}, {Munn},
  {Narayanan}, {Nash}, {Neilsen}, {Neswold}, {Newberg}, {Nichol}, {Nicinski},
  {Nonino}, {Okada}, {Okamura}, {Ostriker}, {Owen}, {Pauls}, {Peoples},
  {Peterson}, {Petravick}, {Pier}, {Pope}, {Pordes}, {Prosapio},
  {Rechenmacher}, {Quinn}, {Richards}, {Richmond}, {Rivetta}, {Rockosi},
  {Ruthmansdorfer}, {Sandford}, {Schlegel}, {Schneider}, {Sekiguchi}, {Sergey},
  {Shimasaku}, {Siegmund}, {Smee}, {Smith}, {Snedden}, {Stone}, {Stoughton},
  {Strauss}, {Stubbs}, {SubbaRao}, {Szalay}, {Szapudi}, {Szokoly}, {Thakar},
  {Tremonti}, {Tucker}, {Uomoto}, {Vanden Berk}, {Vogeley}, {Waddell}, {Wang},
  {Watanabe}, {Weinberg}, {Yanny}, {Yasuda}, \& {SDSS
  Collaboration}}]{York2000}
{York} D.~G. {et~al.}, 2000, \aj, 120, 1579

\end{thebibliography}

\appendix

\section{The photometric redshift tables in SDSS DR12}
\label{sec:appendix}

Here we give a description of each column in the published tables, either referencing a concept used in the article, or detailing it here. With \texttt{\{ugriz\}} we denote that there is a column for each of the five SDSS \texttt{ugriz} broad-band magnitudes, with the single corresponding (capitalized) letter present in the column name.

\subsection{The Photoz table}
\label{sec:appendix1}

\begin{itemize}
	\item \texttt{objID} -- the SDSS \texttt{objID} of the query galaxy.
	\item \texttt{z} -- $z_{\mathrm{phot},i}$ in Eq.\ref{eq:redshift}, i.e. the photometric redshift. It takes the value $-9999$ when there was an error in the fitting algorithm.
	\item \texttt{zErr} -- $\delta z_{\mathrm{phot},i}$ in Eq.\ref{eq:redshifterror}, i.e. the photometric redshift error estimate. It takes the value $-9999$ when there was an error in the fitting algorithm.
	\item \texttt{nnCount} -- $l$, the number of nearest neighbours used in the local linear regression, with outliers excluded from the total of $k=100$, as described in Sec.~\ref{sec:linreg}. It takes the value $-9999$ when there was an error in the fitting algorithm.
	\item \texttt{nnVol} -- the volume of the bounding box of the $k=100$ nearest neighbours.
	\item \texttt{photoErrorClass} -- the photometric error class described in Sec.~\ref{sec:errordiscussion}, Tab.~\ref{tab:errorclasses1} and Tab.~\ref{tab:errorclasses2}.
	\item \texttt{nnObjID} -- the SDSS \texttt{objID} of the first nearest neighbour.
	\item \texttt{nnSpecz} -- the spectroscopic redshift ($z_{\mathrm{spec}}$) of the first nearest neighbour.
	\item \texttt{nnFarObjID} -- the SDSS \texttt{objID} of the farthest, 100th nearest neighbour.
	\item \texttt{nnAvgZ} -- the average redshift of the $k=100$ nearest neighbours.
	\item \texttt{distMod} -- the distance modulus ($DM_i$) corresponding to \texttt{z}, if available, or \texttt{nnAvgZ}. See the end of Sec.~\ref{sec:intro} for the adopted cosmology.
	\item \texttt{lumDist} -- the luminosity distance in $Mpc$ corresponding to \texttt{z}, if available, or \texttt{nnAvgZ}. See the end of Sec.~\ref{sec:intro} for the adopted cosmology.
	\item \texttt{chisq} -- the $\chi^2$ value of the spectral template fit, i.e. $\chi^2 = \sum_{p = 1}^{D} \left(\frac{ m_{p,i} - \left(s_p\left(z=z_i,t=t_i\right) - m_{0,i}\right)}{\Delta m_{p,i}}\right)^2$, using the notation of Eq.~\ref{eq:templatefit}.
	\item \texttt{rnorm} -- the residual Euclidean norm of the spectral template fit, i.e. $\left(\sum_{p = 1}^{D} \left(m_{p,i} - \left(s_p\left(z=z_i,t=t_i\right) - m_{0,i}\right)\right)^2\right)^{0.5}$, using the notation of Eq.~\ref{eq:templatefit}.
	\item \texttt{bestFitTemplateID} -- $t_i$, the identifier of the best-fitting spectral template. See Tab.~\ref{tab:templatelist} for the corresponding names in \citet{Dobos2012}.
	\item \texttt{synth\{ugriz\}} -- the synthetic magnitude of the best-fitting spectral template, i.e. $s_p\left(z=z_i,t=t_i\right) - m_{0,i}$, using the notation of Eq.~\ref{eq:templatefit}.
	\item \texttt{kcorr\{ugriz\}} -- the $K$-correction to $z=0$, i.e. $K_{p,i}\left(z=0\right) = s_p\left(z=z_i,t=t_i\right)-s_p\left(z=0,t=t_i\right)$, using the notation of Eq.~\ref{eq:templatefit}.
	\item \texttt{kcorr\{ugriz\}01} -- the $K$-correction to $z=0.1$, i.e. $K_{p,i}\left(z=0.1\right) = s_p\left(z=z_i,t=t_i\right)-s_p\left(z=0.1,t=t_i\right)$, using the notation of Eq.~\ref{eq:templatefit}.
	\item \texttt{absMag\{ugriz\}} -- the rest-frame absolute magnitude of the galaxy, i.e. $m_{p,i}-K_{p,i}\left(z=0\right)-DM_i$, using the notation of Eq.~\ref{eq:templatefit}.
\end{itemize}

\begin{table}
	\begin{tabular}{c | c | c | c | c | c}
		\hline
		
		$t_i$ & Name & $t_i$ & Name & $t_i$ & Name \\
		
		\hline
		
	
		1 & Red P & 15 & Blue P & 29 & RED 1 \\
		2 & Red H$\alpha$ & 16 & Blue H$\alpha$ & 30 & RED 2 \\
		3 & Red SF & 17 & Blue SF & 31 & RED 3 \\
		4 & Red A+\ion{H}{ii} & 18 & Blue A+\ion{H}{ii} & 32 & RED 4 \\
		5 & Red L & 19 & Blue L & 33 & RED 5 \\
		6 & Red S & 20 & Blue S & 34 & SF 1 \\
		7 & Red all & 21 & Blue all & 35 & SF 2 \\
		8 & Green P & 22 & All P & 36 & SF 3 \\
		9 & Green H$\alpha$ & 23 & All H$\alpha$ & 37 & SF 4 \\
		10 & Green SF & 24 & All SF & 38 & SF 5 \\
		11 & Green A+\ion{H}{ii} & 25 & All A+\ion{H}{ii} &  & \\
		12 & Green L & 26 & All L &  & \\
		13 & Green S & 27 & All S &  & \\
		14 & Green all & 28 & All all &  & \\

		\hline
	\end{tabular}
	\caption{The name in \citet{Dobos2012} that corresponds to the $t_i$ (or \texttt{bestFitTemplateID}) template identifier used in this article.}
	\label{tab:templatelist}
\end{table}

\subsection{The PhotozErrorMap table}

\begin{itemize}
	\item \texttt{CellID} -- The unique identifier of the cell in the grid. The grid spans the $r$-band magnitude, and the $g-r$, $r-i$ colours.
	\item \texttt{rMag} -- The centerpoint of the cell in $r$-band magnitude. Linear size of a cell: $0.5$.
	\item \texttt{gMag\_Minus\_rMag} -- The centerpoint of the cell in $g-r$ colour. Linear size of a cell: $0.01$.
	\item \texttt{rMag\_Minus\_iMag} -- The centerpoint of the cell in $r-i$ colour. Linear size of a cell: $0.01$.
	\item \texttt{countInCell} -- The number of training set galaxies within the cell (denoted below with $N$).
	\item \texttt{avgPhotoZ} -- The average photometric redshift of training set galaxies in the cell, i.e. $\frac{\sum_{i=1}^N z_{\mathrm{phot},i}}{N}$, using the notation of Sec.~\ref{sec:linreg}.
	\item \texttt{avgSpectroZ} -- The average spectroscopic redshift of training set galaxies in the cell, i.e. $\frac{\sum_{i=1}^N z_{\mathrm{spec},i}}{N}$, using the notation of Sec.~\ref{sec:linreg}.
	\item \texttt{avgRMS} -- The rms of the redshift estimation error for training set galaxies in the cell, i.e. $\left(\frac{\sum_{i=1}^N \left(z_{\mathrm{phot},i}-z_{\mathrm{spec},i}\right)^2}{N}\right)^{0.5}$, using the notation of Sec.~\ref{sec:linreg}.
	\item \texttt{avgEstimatedError} -- The average redshift error estimate for training set galaxies in the cell, i.e. $\frac{\sum_{i=1}^N \delta z_{\mathrm{phot},i}}{N}$, using the notation of Sec.~\ref{sec:linreg}.
	\item \texttt{avgNeighborZStDev} -- The average standard deviation of the redshifts of the $k=100$ nearest neighbours, for every training set galaxy in the cell. Denoting the standard deviation of the $z_{\mathrm{spec}}$ of the neighbours with $\sigma_i \left(z_{\nn}\right)$, it is $\frac{\sum_{i=1}^N \sigma_i \left(z_{\nn}\right)}{N}$.
\end{itemize}

\bsp

\label{lastpage}

\end{document}